\documentclass[useAMS,usenatbib]{mnras}
\usepackage[british]{babel}             
\usepackage{newtxtext}               
\usepackage[slantedGreek]{newtxmath}    
\let\la=\lesssim 
\usepackage[T1]{fontenc}               
\usepackage{graphicx}

\graphicspath{./{Figs/}}
\DeclareGraphicsExtensions{.pdf}

\title[An ultra-dense fast outflow in a quasar at $z=2.4$]{An ultra-dense fast outflow in a quasar at $\mathbf{ z=2.4}$}

\author[R.~J.\ Williams et al.]{R.~J.\ Williams$^{1,}$$^{2}$, R.\ Maiolino$^{1,}$$^{2}$, Y.\ Krongold$^{3}$, S.\ Carniani$^{1,}$$^{2,}$$^{4,}$$^{5}$, G.\ Cresci$^{5}$,  
 \newauthor F.\ Mannucci$^{5}$, A.\ Marconi$^{5}$\\ \\
$^1$ Cavendish Laboratory, University of Cambridge, 19 J.J. Thomson Ave., Cambridge\\
$^2$ Kavli Institute for Cosmology, University of Cambridge, Madingley Road, Cambridge\\
$^3$ Instituto de Astronomia, Universidad Nacional Autonoma de Mexico, Apartado Postal 70-264, 04510 Mexico DF, Mexico\\
$^4$ Dipartimento di Fisica e Astronomia, Universit\`a di Firenze, Via G. Sansone 1, 1-50019, Sesto Fiorentino (Firenze), Italy\\
$^5$ INAF - Osservatorio Astrofisico di Arcetri, Largo E. Fermi 5, 50125 Firenze, Italy\\
}

\begin{document}

	\date{Accepted XXX. Received XXX}
	\pagerange{\pageref{firstpage}--\pageref{fig:3epochs}} \pubyear{2016}
	\maketitle
	\label{firstpage}

\begin{abstract}
We present Adaptive Optics assisted near-IR integral field spectroscopic observations of a luminous quasar at $z = 2.4$, previously observed as the first known example at high redshift
of large scale quasar-driven outflow quenching star formation in its host galaxy. The nuclear spectrum shows broad and blueshifted
H$\upbeta$ in absorption, which is tracing outflowing gas with high densities ($>10^8$ -- $10^9$~cm$^{-3}$) and velocities in
excess of 10,000~km~s$^{-1}$. The properties of the outflowing clouds (covering factor, density, column density and inferred
location) indicate that they likely originate from the Broad Line Region. The energetics of such nuclear regions is consistent
with that observed in the large scale outflow, supporting models in which quasar driven outflows originate from the nuclear region
and are energy conserving. We note that the asymmetric profile of both the H$\upbeta$ and H$\upalpha$ emission lines is likely due
to absorption by the dense outflowing gas along the line of sight. This outflow-induced asymmetry has implications on the
estimation of the black hole mass using virial estimators, and warns about such effects for several other quasars characterized by
similar line asymmetries. More generally, our findings may suggest a broader revision of the decomposition and interpretation of
quasar spectral features, in order to take into account the presence of potential broad blueshifted Balmer absorption lines. Our
high spatial resolution data also reveal  redshifted nebular emission lines, which could be potentially
tracing an inflowing stream.

\end{abstract}



\begin{figure*}
	\centering
	\includegraphics[width=0.88\textwidth]{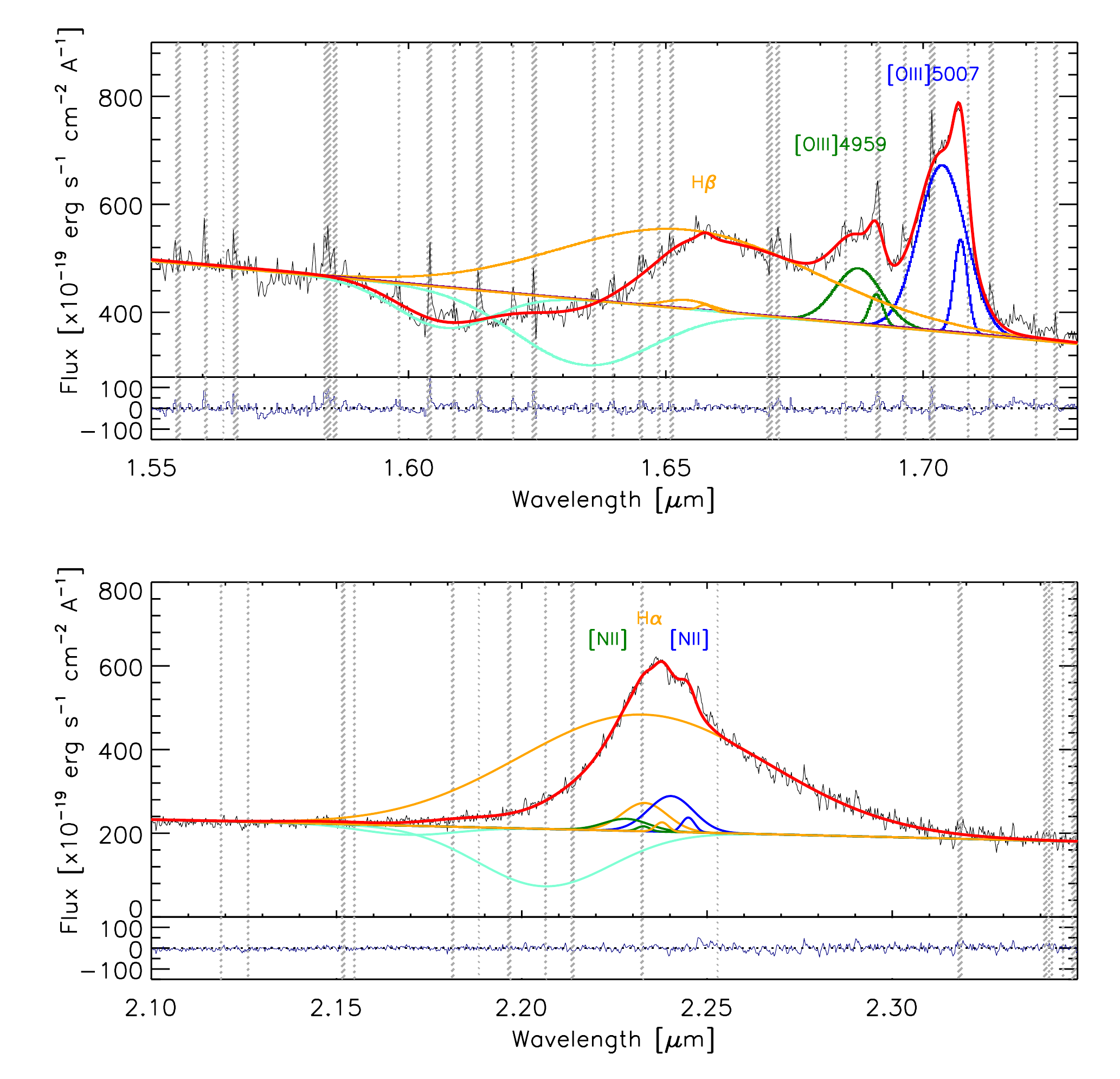}
	\caption{Spectrum extracted from the central square arcsec, showing also the results from simultaneously fitting the \textit{H} and \textit{K}-band emission lines. \textit{Top panel:} \textit{H}-band region of the spectrum, fitting [\ion{O}{iii}]$\lambda$5007 (blue), [\ion{O}{iii}]$\lambda$4959 (green), H$\upbeta$ emission (orange) and H$\upbeta$ absorption (aqua). \textit{Bottom panel:} \textit{K}-band region of the spectrum, fitting H$\upalpha$ (orange), [\ion{N}{ii}]$\lambda$6584 (blue), [\ion{N}{ii}]$\lambda$6548 emission (green) and H$\upalpha$ absorption (aqua). In both spectra the total resulting fit is shown by the red line and the residuals from the fit are shown underneath each. The grey dashed lines indicate the spectral regions affected by strong OH lines.}
	\label{fig:fitting}
\end{figure*}

\section{Introduction}
Quasar feedback is thought to be an important mechanism in regulating the growth and evolution of galaxies. The number density of galaxies falls off steeply at the high mass end, which suggests that some process stops galaxies from growing too massive. One favoured explanation is outflows driven by QSOs, which are thought to expel gas out of galaxies and hence quench star formation \cite[although `starvation/strangulation' has been recently proposed as an alternative quenching mechanism, e.g.][]{peng15}. There have been many examples of QSO-driven outflows observed in local galaxies \cite[e.g.][]{feruglio10,sturm11,cicone14}, however QSO feedback is expected to be even more prevalent at higher redshifts ($z\approx 2$), as this is around the peak of galaxies star formation and black holes growth, with the level of star formation decreasing ever since. In the recent years there have been many detections of QSO-driven outflows also at high-$z$, in ionized, atomic and molecular gas phases \citep[e.g.][]{harrison12,borguet12,maiolino_sep12,cicone15,carniani15} including observations of actual feedback on to the host galaxy \citep[e.g.][]{canodiaz11,cresci15,ebrero15, carniani16}. 

Here we present new VLT-SINFONI near-IR integral field observations using adaptive optics (AO) of the quasar 2QZJ002830.4--281706, at $z = 2.401$. This object was previously observed with SINFONI in 2006, originally selected from a sample of strong [\ion{O}{iii}]$\lambda$5007 emitters \citep{shemmer04}. The forbidden [\ion{O}{iii}]$\lambda$5007 emission line provides a good tracer of outflowing ionized gas. This transition has a critical density much lower than that found in the Broad Line Regions (BLR) of AGNs, hence its profile is not contaminated by emission from high velocity clouds in the BLR. In the original observations, \cite{canodiaz11} were able to map the kinematics of the [\ion{O}{iii}]$\lambda$5007 emission line, which revealed massive outflows on kiloparsec scales. In addition, by accounting for the broad H$\upalpha$ line profile originating from the BLR, they identify a narrow H$\upalpha$ component tracing star formation in the host galaxy. Interestingly this narrow H$\upalpha$ component is not uniformly distributed across the galaxy, but instead lies mostly in regions not affected by a strong outflow, which provided the first evidence that outflows from QSOs can suppress star formation. This result has been recently confirmed by \cite{carniani15} through additional SINFONI observations. Here we present new SINFONI observations of this galaxy, assisted by AO, which enable us to obtain a more detailed picture of the physical processes in the central region of this quasar.

\begin{table*}
	\caption{The best-fitting parameters from the fitting of the nuclear region for each component, where we give the central wavelength, the total flux, FWHM and the velocity relative to the peak of the [\ion{O}{iii}]$\lambda$5007 at zero velocity. Note that the quoted flux errors only refer to the relative flux uncertainties within each of the two bands; however, an additional systemic uncertainty of about 50 per cent should be added to the data in each of the two bands to account for absolute calibration accuracy (see text).}
	\label{tab:fitpars} 
	\begin{tabular}{@{}lccccc@{}}  
	\hline
	Line & Component & $\lambda$  & Flux ($\times10^{-16}$ & FWHM & Vel. \\
	& &($\upmu$m)& erg s$^{-1}$ cm$^{-2})$ & (km~s$^{-1}$) &  (km~s$^{-1}$)\\
	\hline
	\multicolumn{6}{c}{\textit{H}-band}\\
	$[\ion{O}{iii}]$5007 & Nar & $1.7073$ & $  5.7 \pm   0.4$ & $   231 \pm    11$ & $     0 \pm     8$ \\
	$[\ion{O}{iii}]$5007 & Int & $1.7037$ & $ 35.7 \pm   0.6$ & $   811 \pm     9$ & $  -627 \pm    17$ \\
	$[\ion{O}{iii}]$4959 & Nar & $1.6909$ & $  1.9 \pm   0.4$ & $   231 \pm    11$ & $     0 \pm     8$ \\
	$[\ion{O}{iii}]$4959 & Int & $1.6874$ & $ 11.9 \pm   0.6$ & $   811 \pm     9$ & $  -627 \pm    17$ \\
	$\rm H\upbeta$ & Nar & $1.6577$ & $ 0.35 \pm  0.10$ & $   231 \pm    11$ & $     0 \pm     8$ \\
	$\rm H\upbeta$ & Int & $1.6542$ & $  1.9 \pm   0.5$ & $   811 \pm     9$ & $  -627 \pm    17$ \\
	$\rm H\upbeta$ & Broad & $1.6539$ & $   91 \pm     4$ & $  4518 \pm    85$ & $  -678 \pm   214$ \\
	$\rm H\upbeta$ Absorp. & Comp.1 & $1.6341$ & $  -40 \pm     4$ & $  2449 \pm   103$ & $ -4256 \pm    58$ \\
	$\rm H\upbeta$ Absorp. & Comp.2 & $1.6070$ & $ -17.7 \pm    1.2$ & $  1759 \pm    83$ & $ -9172 \pm    93$ \\
	\hline
	\multicolumn{6}{c}{\textit{K}-band}\\
	$\rm H\upalpha$ & Nar & $2.2379$ & $  1.0 \pm   0.3$ & $   231 \pm    11$ & $     0 \pm     8$ \\
	$\rm H\upalpha$ & Int & $2.2332$ & $ 10.3 \pm   0.8$ & $   811 \pm     9$ & $  -627 \pm    17$ \\
	$\rm H\upalpha$ & Broad & $2.2328$ & $   235 \pm    11$ & $  4518 \pm    85$ & $  -678 \pm   214$ \\
	$\rm H\upalpha$ Absorp. & Comp.1 & $2.2061$ & $  -61 \pm     9$ & $  2449 \pm   103$ & $ -4256 \pm    58$ \\
	$\rm H\upalpha$ Absorp. & Comp.2 & $2.1694$ & $    -6 \pm      2$ & $  1759 \pm    83$ & $ -9172 \pm    93$ \\
	$[\ion{N}{ii}]$6584 & Nar & $2.2449$ & $  1.5 \pm   0.3$ & $   231 \pm    11$ & $     0 \pm     8$ \\
	$[\ion{N}{ii}]$6584 & Int & $2.2403$ & $ 13.0 \pm   1.1$ & $   811 \pm     9$ & $  -627 \pm    17$ \\
	$[\ion{N}{ii}]$6548 & Nar & $2.2328$ & $ 0.51 \pm  0.10$ & $   231 \pm    11$ & $     0 \pm     8$ \\
	$[\ion{N}{ii}]$6548 & Int & $2.2282$ & $  4.3 \pm   0.3$ & $   811 \pm     9$ & $  -627 \pm    17$ \\
	\hline
	\end{tabular}\\
\end{table*}

\section{Observations and Data reduction}
We used the near-IR integral field spectrometer SINFONI to observe 2QZJ002830.4--281706 (2QZJ0028 hereafter) assisted by its AO module and the laser star facility. Observations were taken over three nights between July and November in 2013 using both the \textit{H} and \textit{K} gratings, delivering a spectral resolution of R = 3000 and R = 4000, respectively. At $z = 2.401$ this allows the H$\upalpha$+[\ion{N}{ii}] lines to be observed in the \textit{K}-band and H$\upbeta$+[\ion{O}{iii}] lines to be observed in the \textit{H}-band. The total on source integration time was 90 minutes on \textit{H}-band and 150 minutes on \textit{K}-band. We adopted the 0.05 arcsec $\times$ 0.1 arcsec pixel scale. Thanks to the AO the PSF during the observations was generally around 0.15 arcsec. 

The data reduction involved the following steps. To begin, we removed cosmic rays from the raw data using the L.A.Cosmic procedure \citep{vandokkum_nov01}. We then used the ESO-SINFONI pipeline (version 2.3.3) to do the flat fielding, wavelength calibration and reconstruct a non-sky subtracted cube for each observation, in which the pixels are resampled to a symmetric angular size of 0.05 arcsec $\times$ 0.05 arcsec. We next implemented the sky subtraction technique from \cite{davies_mar07}  to perform a better removal of the residual OH airglow emission lines from the data. We corrected for the atmospheric absorption and instrumental response using a telluric standard star, which also provides the initial flux calibration, before finally combining all the single OB cubes to produce a final science cube.

However, we note that the absolute flux calibration is very difficult since the Strehl ratio of the AO observations in the two bands and on different dates can vary significantly and is difficult to control (since the AO correction, especially using the Laser star facility, can vary significantly in time), resulting in large uncertainties on the fraction of light lost into the seeing-limited PSF. We have decided to cross-calibrate the two bands by scaling their fluxes to match the \textit{H} and \textit{K}-band magnitudes from the 2MASS All Sky Catalogue of point sources \citep{cutri03},
 which however were taken on a different epoch. As a consequence of these issues, we estimate that the accuracy of  the absolute flux calibration of each of the two bands is about 50 per cent.

\section{Analysis of nuclear spectrum}
\label{nuclear}

Fig.~\ref{fig:fitting} shows the spectrum extracted from the central square arcsecond (which includes most of the light in our data). We clearly detect strong H$\upalpha$ emission in the \textit{K}-band and H$\upbeta$ and the two [\ion{O}{iii}] lines in the \textit{H}-band. H$\upalpha$ and H$\upbeta$ are clearly dominated by emission from the BLR. [\ion{O}{iii}] is also clearly broad and strongly asymmetric, skewed towards the blue, tracing the quasar-driven outflow already identified by previous studies. However, [\ion{O}{iii}] also shows evidence for a narrow component, tracing dynamically quiescent gas in the host galaxy which was also found by previous studies. In order to disentangle the different emission components, we fit the \textit{H} and \textit{K}-band spectra with a combination of Gaussians each describing a different component of the emission. 

First, we follow \cite{canodiaz11} and \cite{carniani15} by fitting the prominent [\ion{O}{iii}]$\lambda$5007 emission line using two Gaussian components with two different widths, which we call `narrow' ($\sigma<$ 300 km~s$^{-1}$, mostly tracing  the dynamically `quiescent' gas) and `intermediate' (300 $< \sigma<$ 1500 km~s$^{-1}$, tracing the outflow). The same components (linked in velocity and velocity width to [\ion{O}{iii}]$\lambda$5007) are used for the second [\ion{O}{iii}]$\lambda$4959 line and later also for the much weaker [\ion{N}{ii}]$\lambda$6584 and [\ion{N}{ii}]$\lambda$6548 lines. The doublet flux ratios $F$([\ion{O}{iii}]$\lambda$5007)/$F$([\ion{O}{iii}]$\lambda$4959) and $F$([\ion{N}{ii}]$\lambda$6584)/$F$([\ion{N}{ii}]$\lambda$6548) are also linked to their relative Einstein spontaneous transition coefficients. The continuum is fitted with a simple power law. We also attempted to include in the fit the potential \ion{Fe}{ii} emission `humps' by using the  ion Fe {\sc ii} template provided by \cite{Tsuzuki06}, however we found that the spectrum does not require this contribution, which is set to zero by the fit. A more extensive discussion about the lack of \ion{Fe}{ii} emission contributing to the spectrum is given in Appendix~\ref{sect:feii}. In addition, we attempt to also include [\ion{S}{ii}]$\lambda 6717,6731$ in the fit, however the inferred components are not significant and not required by the fit; the upper limit on the flux of the doublet implies $[\ion{S}{ii}]/[\ion{N}{ii}]<0.75$ for the narrow component and $[\ion{S}{ii}]/[\ion{N}{ii}]<0.2$ for the intermediate component, which are both consistent with the broad spread of [\ion{S}{ii}]/[\ion{N}{ii}] typically observed in the NLR of AGNs \citep{stern13}.

The resulting best-fitting parameters and associated errors for all components of the nuclear fit are shown in Table~\ref{tab:fitpars}.

\subsection{Broad Balmer absorption: evidence for an ultra-dense outflow}

The H$\upbeta$ profile is extremely intriguing. It is clearly much broader than the [\ion{O}{iii}] lines, and clearly dominated by emission from the BLR. The profile is also asymmetric, with the blue shoulder declining more steeply than the red shoulder. However, the most interesting feature is that there is a clear broad absorption feature blue ward of the line. We checked that such absorption is not an artefact associated with imperfect atmospheric absorption removal or associated with the telluric calibration star (the feature is not located in a region of bad atmospheric absorption and it persists when using different telluric standard stars, or even without correcting for telluric absorption at all, i.e. without using any telluric standard star). We also checked that it is not an artefact associated with the flat fielding or with the background subtraction (the feature persists even when using different flats, different background subtraction schemes, and is absent in the flux calibration star and in the PSF calibration star), therefore this is a real feature associated with the quasar. There are no prominent \ion{Fe}{II} resonant absorption lines that could explain this absorption \citep{shi16}.

The most straightforward interpretation is that this feature is tracing H$\upbeta$ absorption associated with a dense high velocity outflow along our line of sight. Balmer lines absorption requires a significant population of the $n=2$ hydrogen level. Since the latter level has an extremely short lifetime for radiative transition to the fundamental level, the only way to keep it populated is through collisional excitation in an extremely dense medium and with temperatures of the order of $10^4$ K (required to thermally populate the level, while temperatures significantly higher than $\sim 2\times 10^4$~K leads to significant ionization of the medium), or it must originate from a region with very high optical thickness which keeps the $n=2$ level excited \cite[e.g.][]{hall07,ji12}, which also implies very high densities. We will see more quantitatively later in this section that any of these excitation mechanisms require a very high density in the clouds, larger than about $\rm 10^9~ cm^{-3}$.

H$\upbeta$ absorption is commonly observed in Type II supernova ejecta \cite[e.g.][]{Kleiser11}. Detections of H$\upbeta$ absorption have also been reported in a dozen of Seyferts and QSOs spectra \cite[e.g][]{Hutchings02,hall02,aoki06,hall07,wang08,aoki10,shapovalova10,ji12,ji13,wang15,zhang15,shi16} and are generally interpreted in terms of absorption by dense clouds, possibly associated with BLR clouds along the line of sight (which are indeed expected to have densities of about $10^{11}$~cm$^{-3}$ and temperatures of $\sim10^4$~K), although some of these authors have modelled these absorptions on scales slightly larger than the BLR.

However, these previous detections of H$\upbeta$ absorption in AGN were characterized by a much narrower width (FWHM $ \sim 200$ to $1000$~km~s$^{-1}$). The only previous detection of very broad Balmer absorption was found by \cite{zhang15}, in a BAL quasar at $z=0.6$. The strongly blueshifted Balmer absorption in 2QZJ0028 is very broad (width $\sim 10,000$~km~s$^{-1}$). It is in some aspects similar to that observed by \cite{zhang15}, but 2QZJ0028 is at a much higher redshift (around the peak of QSO cosmic luminosity density) and much more luminous ($\lambda L_{5100}=3\times 10^{46}$~erg~s$^{-1}$, while the quasar studied by \cite{zhang15} is 30 times less luminous).

In summary, the H$\upbeta$ absorption seen in 2QZJ0028 must be tracing an ultra-dense fast outflow ($v\sim10,000~\rm km~s^{-1}$) along our line of sight.

\subsection{Simultaneous fit with H\boldmath{$\upalpha$}}
\label{simult_fitting}
We do not see any obvious signature of such an absorption feature in H$\upalpha$. The latter has been seen in other studies of AGNs \citep[e.g.][]{wang15,zhang15}. However, for many other previous detections of H$\upbeta$ absorption in AGNs, the corresponding H$\upalpha$ absorption was also not detected \cite[e.g][]{Hutchings02}, and is most likely due to the dominance of the strong H$\upalpha$ in emission. The latter is also the reason why in supernova ejecta the H$\upalpha$ absorption is much shallower than the H$\upbeta$ absorption and sometimes even undetected.

However, the H$\upalpha$ broad component of 2QJZ0028 shows a very asymmetric profile, where the asymmetry is even more pronounced than in the H$\upbeta$ profile, and which is likely indicative of an underlying absorption feature on the blue side of the line. Indeed, the global profile of both H$\upbeta$ and H$\upalpha$ can very nicely be reproduced by a broad  ($\rm FWHM \sim 4500~km~s^{-1}$) symmetric emission component (in addition to the intermediate and narrow component used to fit the [\ion{O}{iii}] lines) and a blueshifted absorption feature, which is fitted with two negative Gaussians. The velocities and widths of the broad Gaussians (both in emission and absorption) are tied to be the same for H$\upalpha$ and H$\upbeta$, only the H$\upbeta$ to H$\upalpha$ flux ratios are left free to vary. The resulting fit is shown in Fig.~\ref{fig:fitting}, along with the residuals and the best fit parameters for these components are given in Table~\ref{tab:fitpars}, illustrating that the whole profile around H$\upbeta$ and H$\upalpha$ are both nicely reproduced.

We have also allowed for the widths of H$\upalpha$ and H$\upbeta$ emission lines to vary independently. The resulting fit does not change significantly. The calculation of the uncertainties on the profile and equivalent widths of the absorption features (later on in this section) will allow for such variation of the relative line widths.

\subsection{Covering factor and constraints on the outflowing clouds size}

Using our absorption line fit, we obtain a resulting rest-frame equivalent width ($EW$) of the H$\upbeta$ absorption of $EW(\rm H\upbeta) = 40 \pm 4~\AA$. According to the relation $N = (m_{\rm e} c^2 W_\lambda)/(\pi e^2 f\lambda_0^2 )$ \cite[][ where $N$ is the column density, $W_\lambda$ denotes the $EW$, $f$ is the oscillator strength and $\lambda _0$ is the rest frame wavelength]{jenkins86}, which applies to the linear part of the curve of growth, this $EW$ would imply a column density of hydrogen atoms whose $n=2$ energy level is populated as $N_{\rmn{H}_0,n=2} \approx (1.60 \pm 0.14) \times 10^{15}$ cm$^{-2}$. However, we will shortly see that the hydrogen absorption lines must be saturated, therefore this column density is actually a conservative lower limit to the column density of hydrogen atoms whose $n=2$ level is populated. The latter also translates into a much larger column of neutral hydrogen, since in ionized clouds in the vicinity of AGNs the fraction of neutral hydrogen is generally a small fraction of the ionized hydrogen, as we will see more quantitatively later in the next subsection.

\begin{figure}
	\centering
	\includegraphics[width=0.48\textwidth]{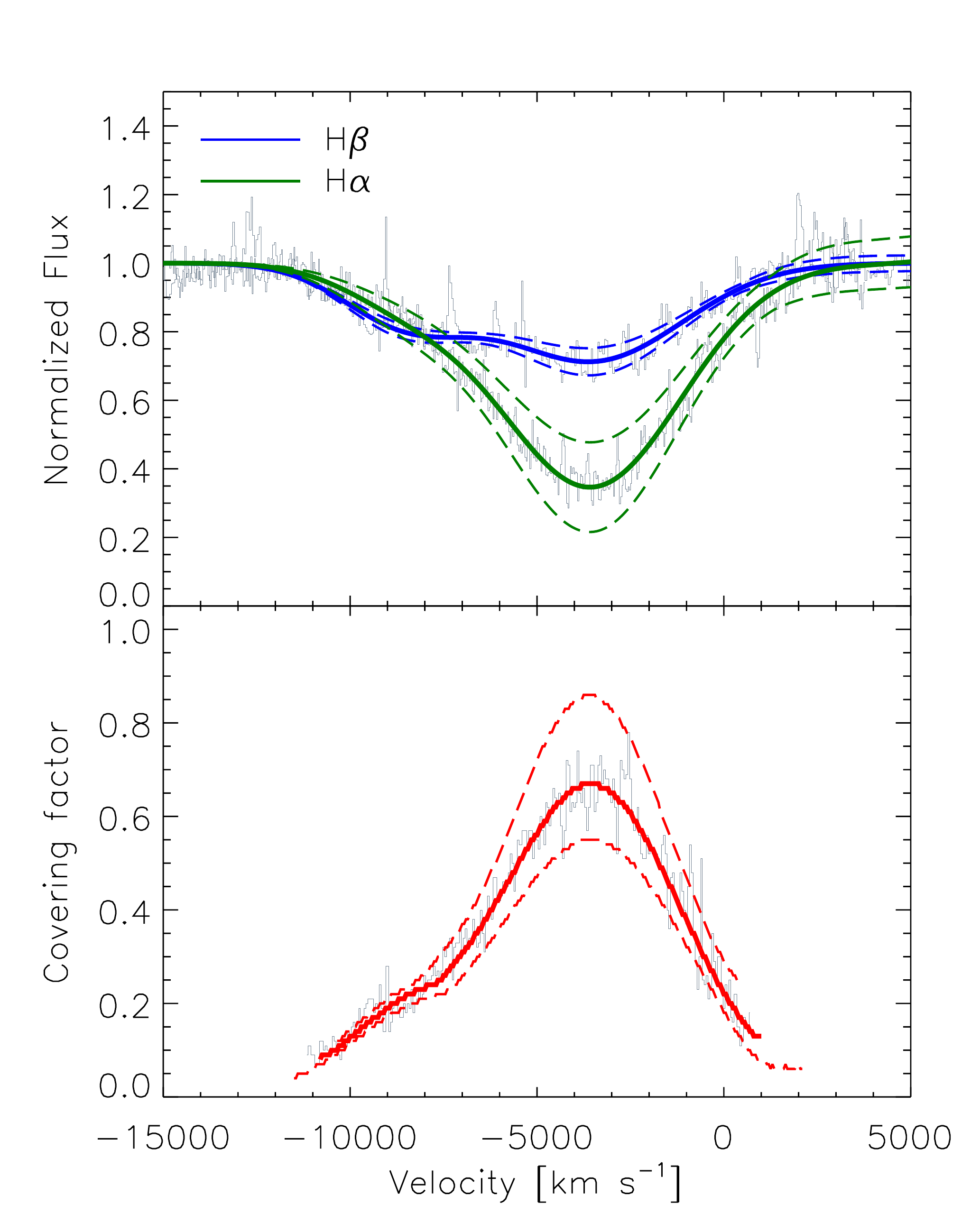}
	\caption{\textit{TOP}: Comparison of the H$\upalpha$ (green) and H$\upbeta$ (blue) absorption line fits after normalizing by the continuum. \textit{BOTTOM}: Derived covering factor as a function of velocity. Note that in both panels, the velocity scale is set to the peak of the [\ion{O}{iii}]$\lambda$5007 at zero velocity.}
	\label{fig:absorp_prof}
\end{figure}

It is also interesting to compare the depths of the H$\upbeta$ and H$\upalpha$ absorptions inferred from the line profiles. The top panel in Fig.~\ref{fig:absorp_prof} shows the velocity profile of the two absorption features. The black line shows the continuum normalized spectra in which the emission components have been removed, while the coloured solid lines show the results of the Gaussian fits discussed in the previous section (the dashed lines show the associated uncertainties).

The equivalent width of H$\upalpha$ is $EW(\rm{H\upalpha}) = 96 \pm 16~\AA$, which gives a ratio of $EW(\rm H\upalpha$)/$EW(\rm H\upbeta) = 2.4$, much lower than expected in the case of unsaturated absorption, i.e. $(f_{\rm H\upalpha}\lambda_{\rm H\upalpha})/(f_{\rm H\upbeta}\lambda_{\rm H\upbeta}) = 7.26$ (where $f_i$ is the oscillator strength of transition $i$ and $\lambda_i$ is the rest wavelength), and which indicates that the lines are saturated. Yet, the H$\upalpha$ absorption line does not extend to zero. This is a clear indication of partial covering, i.e. the
absorbing clouds do not cover the entirety of the QSO optical continuum emission, i.e. the accretion disc . The covering factor ($C_{\rm f}$) can be inferred through the equation $\frac{F_{\rm H\upalpha}(v) - 1}{C_{\rm f}(v)}+1 = \Big( \frac{F_{\rm H\upbeta}(v) - 1}{C_{\rm f}(v)}+1 \Big) ^{r}$, where $r = \frac{f_{\rm H\upalpha}\lambda_{\rm H\upalpha}}{f_{\rm H\upbeta}\lambda_{\rm H\upbeta}}$ and $F_i(v)$ is the continuum level, relative to normalized continuum level outside the through, observed at the velocity $v$ for the transition $i$. By using the H$\upalpha$ and H$\upbeta$ absorption profiles we can derive the covering factor of the outflowing clouds towards the background optical continuum produced by the accretion disc, as a function of the clouds velocity. This is shown in the bottom panel of Fig.~\ref{fig:absorp_prof}. The dashed lines show the errors on the covering factor. The covering factor is always below unity, which implies that the outflowing clouds have a size smaller or comparable (in projection) with the accretion disc. For quasars with black hole mass similar to 2QZJ0028 (see next sections), the accretion disc is expected to have a size of about $0.2$~pc \citep{morgan10}, for the region emitting in the optical, hence this is approximatively the upper limit on the transverse size of the outflowing clouds. The latter is a very conservative upper limit on the clouds, since the outflow is likely composed of multiple clouds, cumulatively producing the observed covering factor. On the high velocity tail ($v\sim 10,000$~km~s$^{-1}$, where we have the tightest constraints because the Balmer absorption is least blended with emission) the covering factor approaches 0.1 or less, hence for the fastest clouds the upper limit on the transverse size is about $0.2~\sqrt{0.1}\approx 0.06$~pc, which is fully consistent with the constraints on the size of the BLR clouds inferred from X-ray occultations in nearby AGNs, i.e. $\sim 10^{-6}$~pc \citep{maiolino10,risalti11}.

The variation of the covering factor as a function of velocity can provide some clues on the nature and origin of the dense outflowing clouds. The steadily decreasing covering factor at velocities higher than about 4,000~km~s$^{-1}$, fits very nicely in the scenario in which the outflowing clouds develop from instabilities in the accretion disc and are then accelerated radially by radiation pressure \citep[e.g.][]{proga00,proga04}; within this scenario the more the clouds are accelerated (to higher velocity), the farther they are from the accretion disc in any radial direction and, as a consequence, the fewer of them are seen in projection against the accretion disc light. The drop in covering factor from $v\sim4,000~$km~s$^{-1}$ to zero velocity can be explained in terms of an increasing contribution to the `background' light from the Balmer {\it emission} lines from the BLR; the size of the BLR is considerably larger than the size of the accretion disc, hence the covering factor of the outflowing clouds at low velocities must decrease accordingly. The latter scenario implies that the outflowing clouds are distributed on a region that is comparable in size to the BLR, i.e. about 1~pc for a quasar with the luminosity of 2QZJ0028 \citep{greene05}.

\subsection{Physical conditions of the outflowing clouds}
\begin{figure}
	\centering
	\includegraphics[width=0.45\textwidth]{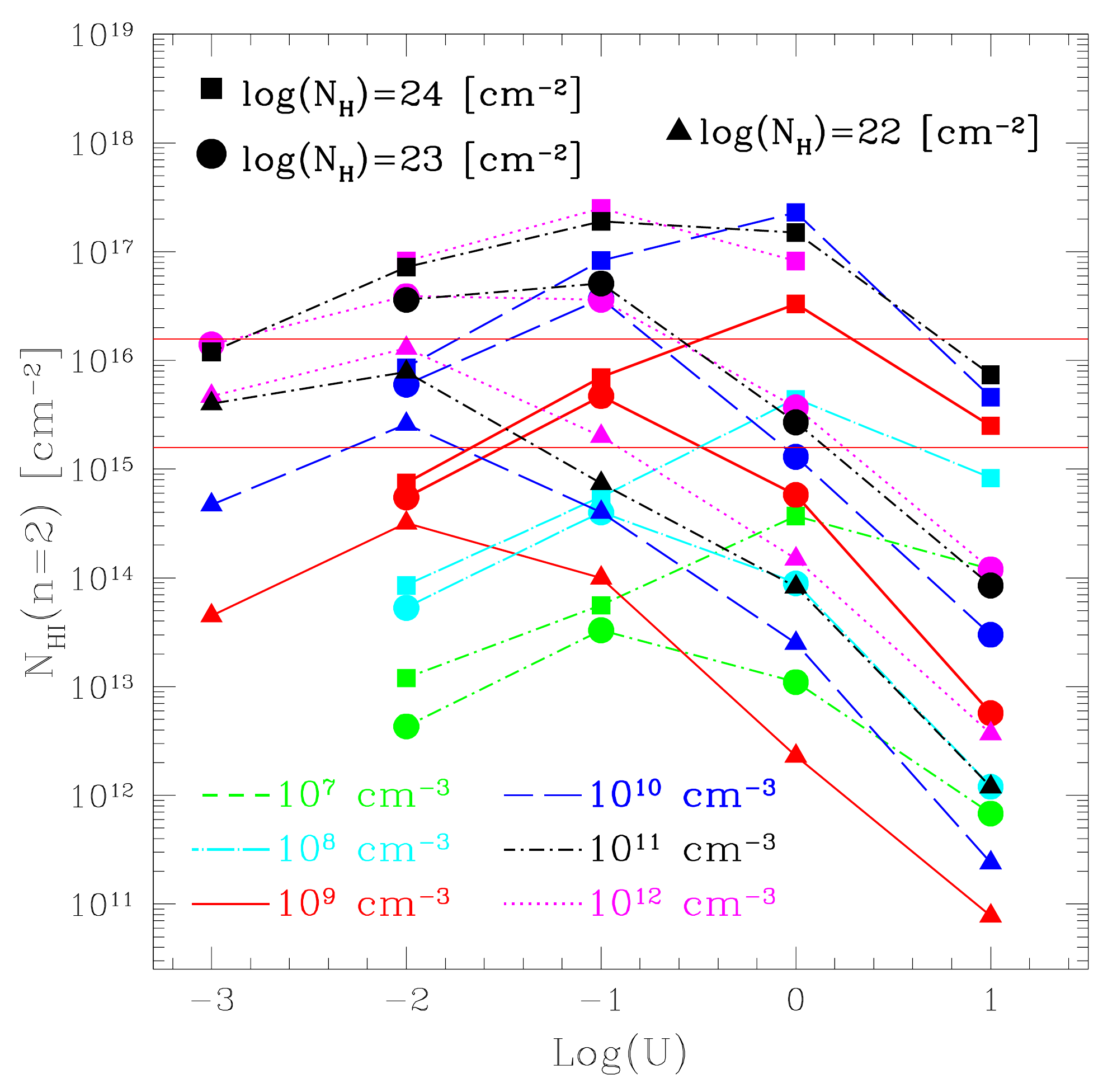}\\
	\caption{Column density of hydrogen atoms whose $n=2$ level is populated ($N_{\rmn{H}_0,n=2}$), as a function of ionization parameter, for a grid of AGN photoionization models in which gas density and total hydrogen column density are varied. The two horizontal red lines embrace the range of $N_{\rmn{H}_0,n=2}$ inferred by the Balmer absorption lines in 2QZJ0028. Note that high densities (generally $ n \rm> 10^9~cm^{-3}$) are needed to account for the observed Balmer absorption.}
	\label{fig:model}
\end{figure}

We have developed simple photoionization models to investigate the physical conditions of the clouds required to have H$\upbeta$ in absorption. More specifically, by adopting a typical AGN Spectral Energy Distribution (the details of the latter are not critical in this case), we have used the photoionization code \textsc{cloudy} to produce a grid of photoionization models by varying the ionization parameter, gas density of the cloud, total column density of hydrogen atoms, and we have then inferred the column of hydrogen atoms that have the $n=2$ level populated (by including all possible population mechanisms discussed above). The results of the models are shown in Fig.~\ref{fig:model}.

In order to compare with the observations, we shall recall that the column density of hydrogen atoms $N_{\rmn{H}_0,n=2}$ inferred for 2QZJ0028 from the H$\upbeta$ absorption $EW$ is a conservative lower limit since the hydrogen absorption lines are saturated. Given that the covering factor is in the range $\sim0.1\--0.7$, the actual column density $N_{\rmn{H}_0,n=2}$ is probably about of the order of a few times, up to an order of magnitude, higher than that inferred from the $EW$(H$\upbeta$) and from the simple linear relation of the curve of growth (a finer determination of the column density would require accessing additional transitions in the UV rest frame). Based on these considerations, the two red horizontal lines embrace the range of $N_{\rmn{H}_0,n=2}$ inferred for 2QZJ0028 (the lower solid horizontal red line showing the conservative lower limit inferred from the H$\upbeta$ absorption $EW$). 

The main result of the comparison of the models with the observational constraints is that large gas densities are required to meet the observational constraints on the H$\upbeta$ absorption. Specifically, densities higher than $\rm 10^9 ~cm^{-3}$ and up to  $\rm 10^{12} ~cm^{-3}$ are generally needed (although one of the models with $n=\rm 10^8 ~cm^{-3}$ can in principle marginally meet the constraints). These are densities typically inferred for the clouds in the BLR.

It is also important to note that the implied column densities of total hydrogen $N_{\rm H}$ (both neutral and ionized) have to be in the range between $\rm \sim 10^{22}$ and $10^{24}~{\rm cm}^{-2}$. Such columns of gas are observed in the BLR of AGNs. Indeed, the X-ray monitoring of individual eclipsing events have identified absorption associated with BLR clouds transiting along our line of sight with a range of column densities spanning from $\rm \sim 10^{22}$ to $\rm 10^{24}~cm^{-2}$ \citep{elvis04,risaliti09,maiolino10}. Our results are also fully consistent with the column densities and gas densities recently reported by \cite{shi16} for another QSO showing Balmer absorption lines, who use photoionization simulations to obtain column densities on the order of $\sim 10^{23}~\rmn{cm}^{-2}$ and a gas density of $10^{9.1}~\rmn{cm}^{-3}$. 

Summarizing, the outflow traced by the H$\upbeta$ absorption is consistent with being associated with BLR clouds in the outflow as predicted by some models \citep[e.g.][]{elvis00}. Moreover, the width of the absorption features implies that we are not observing a single outflowing cloud, but a large population of clouds spanning a large velocity range, implying that the bulk of the BLR must have a significant outflow component, as expected by some models \citep{elvis00,proga00,proga04}.

\subsection{Comparison with previous observations}
\label{sect:compare}
As mentioned above, two additional near-IR spectra of this quasar were obtained in the past with SINFONI, although without AO. In the previous works presenting the previous observations no H$\upbeta$ absorption was reported \citep{canodiaz11,carniani15}. A comparison of the extracted \textit{H}-band spectra from  observations over all three epochs (including ours) is Appendix~\ref{sect:previous}, Fig.~\ref{fig:3epochs}, and illustrates that actually the new spectrum presented in this work is broadly consistent with the spectra taken at previous epochs, and that actually H$\upbeta$ absorption was present also in the observations. There are two main reasons why H$\upbeta$ absorption was not detected in the previous data: 1) previous observations had a significantly lower signal-to-noise (S/N), which made much more difficult to identify the Balmer absorption; our new data have much higher S/N and the use of AO enable us to better isolate the nucleus from spectral features in the host galaxy (narrow and intermediate components); 2) previous works were mostly focused on the [\ion{O}{iii}] emission, and therefore adopted a much narrower band to fit the continuum, which included the H$\upbeta$ absorption itself.

However, we shall also note that the spectra at the three epochs are not exactly identical, some weak variability is detected, both in the emission and absorption components of H$\upbeta$. This is not unexpected if these features generate in, or close to the BLR, which has a sub-parsec size. 

Finally, we shall also mention that the optical UV-rest-frame spectrum of 2QJZ0028, taken by the 2dF survey several years before our observation, does not show any evidence of absorption associated with the resonant UV lines (in particular \ion{C}{iv}, \ion{Si}{iv} and \ion{Mg}{ii}). Therefore optically this quasar was not classified as a BAL QSO. \cite{zhang15} have noted that only about half of previous detections of Balmer absorption lines are associated with BAL QSOs. Models may be developed in which absorption by UV resonant lines are not present together with Balmer absorption lines, however an alternative explanation is that this is due to variability of the outflow or of its physical conditions, or simply transversal motion of the clouds in the outflow, as discussed above.

\subsection{Mass outflow rate and kinetic power} 

It is extremely difficult to infer the mass outflow rate and kinetic power of the central outflow, because most parameters are extremely uncertain. However, it is useful to have an order of magnitude of these quantities for comparison with both the outflow properties on larger scales and with models.

The mass outflow rate can be estimated as $\dot{M} = \langle \rho \rangle ~\Omega r^2~v$, where $\langle \rho \rangle$ is the average density of the outflowing medium at the distance $r$ from the centre, $\Omega$ is the solid opening angle of the outflow, and $v$ is the velocity of the outflow at the distance $r$ from the centre. We can write $\langle \rho \rangle = n_{\rm c}~m_{\rm H}~f$, where $n_{\rm c}$ is the density of the outflowing clouds and $f$ their filling factor, which can be approximated as $ f\approx C_{\rm f}^3$, where $C_{\rm f}$ again denotes the covering factor. Hence $\dot{M} \simeq n_{\rm c}~m_{\rm H}~C_{\rm f}^3~ \Omega r^2~v$. The opening angle $\Omega$ is unknown; however one of the main aims is to compare with the large scale outflow detected in previous seeing-limited observations, of which only the approaching side is observed \citep{carniani15}; the opening angle of the latter is not well constrained either, but can be approximated with $\Omega \sim 1 ~ \rm sr$. We consider the highest velocity part of the outflow traced by the H$\upbeta$ absorption (where the contamination by H$\upbeta$ emission is minimal), i.e. $v\sim 10,000 ~\rm km~s^{-1}$, here the covering factor is about 0.1, hence $f\sim 10^{-3}$. By taking $r$ as the radius of the BLR expected at the luminosity of 2QJZ0028, i.e. $\sim 0.5 ~\rm pc$ \citep{kaspi05} and by assuming $ n_c=10^9~\rm cm^{-3}$ then we obtain a nuclear outflow rate of  $\dot{M}_{\rm nuc} = 60~\rm M_{\odot}~yr^{-1}$.

In terms of kinetic power we obtain $P_{\rm K,nuc}=\frac{1}{2}\dot{M}_{\rm nuc}~v^2 \approx 1.8\times10^{45}~\rm erg~s^{-1}$. This is about 1 per cent of the quasar bolometric luminosity ($L_{\rm AGN}=2\times10^{47}~\rm erg~s^{-1}$), well below the maximum potential kinetic power expected to be driven by a quasar according to models (about 7 per cent of the quasar bolometric luminosity), implying that the quasar certainly has the capability of driving this outflow, even with a low coupling factor. However, the most interesting result is that the kinetic power of the nuclear outflow is comparable to the kinetic power inferred for the large scale, extended outflow by previous observations: $P_{\rm K,ext}\simeq 1.2\times10^{45}~\rm erg~s^{-1}$. Of course these values are subject to large uncertainties: if the outflowing clouds density is larger than $\rm 10^9~cm^{-3}$ then both the nuclear outflow rate and the nuclear kinetic power increase proportionally; regarding the extended outflow, the mass outflow rate and kinetic power are inferred for the ionized component only, while the molecular and neutral atomic components may carry a substantial fraction of the mass and energy of the outflow \citep{carniani15}. Summarizing, this comparison must be taken with care due to the large uncertainties, however it is interesting that the kinetic power of the outflow on nuclear and large scales are comparable, and therefore consistent with energy-driven scenarios.

\subsection{Frequency of ultra-dense outflows traced by Balmer absorption lines}

It is finally worth discussing whether 2QZJ0028 is a rare case, or if extremely dense and fast outflows traced by Balmer absorption lines represent a significant sub-population of luminous quasars. As mentioned above, H$\upbeta$ absorption has been detected in a few additional AGN/QSOs, but only in one other case with velocities comparable to 2QZJ0028. However, it should be noted that obtaining adequate data for identifying H$\upbeta$ or H$\upalpha$ absorption is rare. Indeed, if dense, fast outflows are typical of luminous quasars (as suggested by the preference for BAL QSO to be hosted in luminous systems, \citep{reichard03,hewett03}, then such luminous quasars are mostly found at high redshifts ($z>1$), for which H$\upbeta$ is redshifted into the near-IR, where spectroscopic surveys are still seriously limited and with statistics several orders of magnitude lower than optical quasar spectroscopic surveys (e.g. SDSS or 2QZ). The next generation of near-IR multi-object spectrographs (e.g. PFS at Subaru, MOONS at VLT), as well as space spectroscopic missions (e.g. \textit{EUCLID}), will certainly greatly expand the statistics in terms of near-IR spectroscopy of distant, luminous quasars. 

However, even with the existing optical and near-IR spectra, one should note that the detection of broad H$\upbeta$ and H$\upalpha$ absorption is not trivial for various reasons. First, the detection of these absorption features require spectra with high S/N. A clear example of this issue is 2QZJ0028 itself: as discussed in Sect.~\ref{sect:compare}, previous observations (even if with the same instrument at an $8$~m class telescope) failed to detect H$\upbeta$ absorption as a consequence of their higher S/N. Secondly, Balmer absorption lines may be largely hidden by the Balmer BLR emission lines; red asymmetric broad Balmer profiles, which are actually rather common among quasars, may be a hint of an underlying blueshifted broad absorption component. Finally, the complex nebular spectrum of quasars, and in particular the \ion{Fe}{ii} complex humps, may also seriously hamper the capability of identifying H$\upbeta$ absorption. In Appendix~\ref{sect:feii} we discuss that, in absence of broad-band spectra, H$\upbeta$ absorption may be mistaken for \ion{Fe}{ii} emission. We even speculate that H$\upbeta$ absorption may have actually affected, in many quasars, the shape of the \ion{Fe}{ii} complex and the resulting \ion{Fe}{ii} templates. That said, our result from 2QZJ0028 suggests a re-analysis of existing quasar spectra with a fresh approach, by including the possible presence of Balmer absorption features.

There are also potential, intrinsic, physical reasons why Balmer absorption may be difficult to detect. Indeed, if the outflowing dense clouds occur close the equatorial direction, as assumed by some models to explain the quasar properties \citep{elvis00,maiolino01}, then many of these quasars may be obscured hence difficult to identify or may have their Balmer lines absorbed and not viable to probe putative blueshifted absorption. The variability of such absorption features, as discussed above, may be an additional factor responsible for the lack of detections.

Summarizing, assessing the occurrence of ultra-dense, fast winds in quasars is not yet feasible with the information currently available (not even at the order-of-magnitude level); but future studies and surveys will certainly provide important constraints.

\subsection{Implications for Black Hole mass estimates}

We have shown that the asymmetry of both H$\upbeta$ and H$\upalpha$ is likely due to absorption by dense outflowing gas along our line of sight. As a consequence, the underlying emission component of H$\upalpha$ is significantly broader than inferred by a simple analysis that does not take into account the absorption profile. This issue has some implications on the estimation of the black hole mass based on the virial methods, which make use of the width of the broad emission lines. The effect is even stronger for those virial estimators that make use of both the line width and line luminosity (since also the latter is suppressed). For instance, by using the relation between $M_{\rm BH}$, $L_{\rm H\upalpha}$ and $\rm FWHM_{\rm H\upalpha}$ provided by \citet[]{greene05}, and by taking the observed profile of H$\upalpha$, without any spectral decomposition, one would infer a black hole mass of  $M_{\rm BH,no-dec}=4\times10^9~\rm M_{\odot}$. By subtracting the intermediate component of H$\upalpha$ (which is associated with the large scale outflow and not with the BLR), the H$\upalpha$ flux decreases slightly but the width of the line increases, resulting in a black hole mass of $M_{\rm BH,no-interm.}=6.8 \times10^9~\rm M_{\odot}$. By taking into account that the asymmetry of the line is due to absorption of the dense outflow along the line of sight and by recovering the intrinsic profile of the H$\upalpha$ emission from our best fit, we obtain a black hole mass $M_{\rm BH,real}=1.2 \times 10^{10}~\rm M_{\odot}$. Therefore, the presence of dense outflows, especially along the line of sight, affecting the line profile, can have a significant effect on the black hole mass determination.

As mentioned above, asymmetric broad line profiles are relatively common in quasars and type 1 AGNs in general (although the fraction of such profiles needs to be quantitatively assessed). Even if a clear Balmer absorption line is not readily seen, the asymmetric profile may result from dense outflows as in 2QZJ0028 and as in a few other cases in which Balmer absorption is clearly seen. As a consequence, the black hole mass estimates in this class of quasars likely needs to be revised with a proper analysis and decomposition of the line profile.

Another effect of the outflows is the implication that a large fraction of the BLR would not be in virial equilibrium, which also has implications on the black holes virial estimators. It has already been suggested that in high luminosity AGNs the effect of radiation pressure on to the BLR clouds implies strong deviation from the virial motions and, even if the clouds are still in rotation around the black hole, the net result is an underestimation of the black hole masses through the classical virial methods \citep{marconi08}, since the BLR clouds are subject to a lower effective gravitational force. However, if in addition a significant fraction of the emission line broadening is due to outflowing motions, this could actually go in the opposite direction of overestimating the black hole mass, since classical virial estimators assume that the hole broadening of the line is due to gravitational motions. A detailed analysis of these  effects and the determination of
potential corrections to the black hole mass estimates goes beyond the scope of this paper.

\begin{figure}
	\centering
	\includegraphics[width=0.48\textwidth]{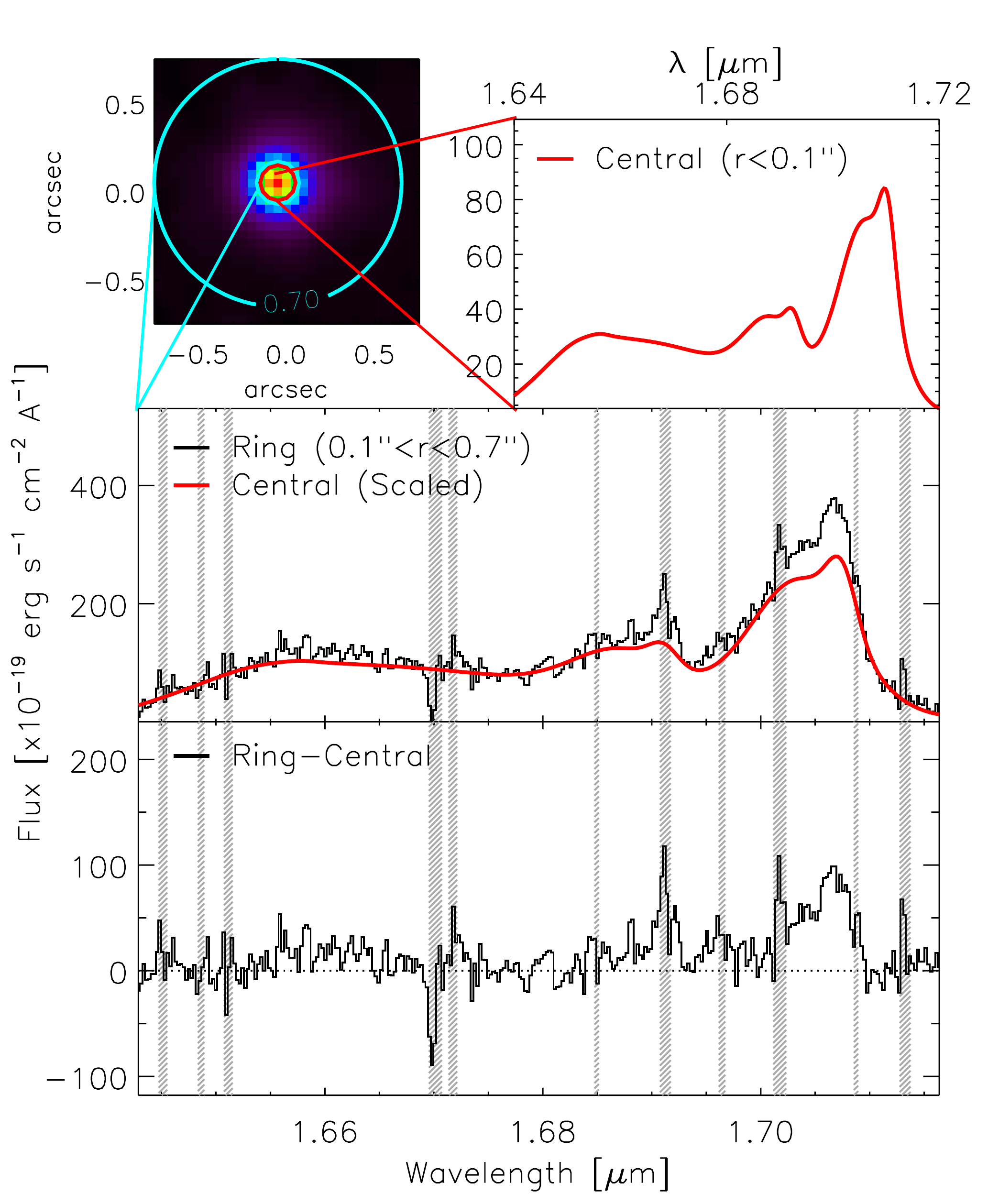}\\
	\caption{Extended emission. \textit{Top, left:} continuum image showing the inner and outer spectral extraction regions. \textit{Top, right:} spectral fit around [\ion{O}{iii}] on the
	spectrum extracted from the very inner region of the QSO ($r<0.1\arcsec$), which is then normalized to the flux of the broad component of the H$\upbeta$ in the spectrum extracted from the outer ring ($0.1\arcsec<r< 0.7\arcsec$) as shown in the \textit{middle} panel. The central spectrum is then subtracted from the outer spectrum and the resulting residual emission is shown in the \textit{bottom} panel, showing indication of extended [\ion{O}{iii}] emission in the outer regions. The grey dashed lines indicate the regions affected by strong OH sky lines.}
	\label{fig:outer_regions}
\end{figure}

\begin{figure}
	\centering
	\includegraphics[width=0.48\textwidth]{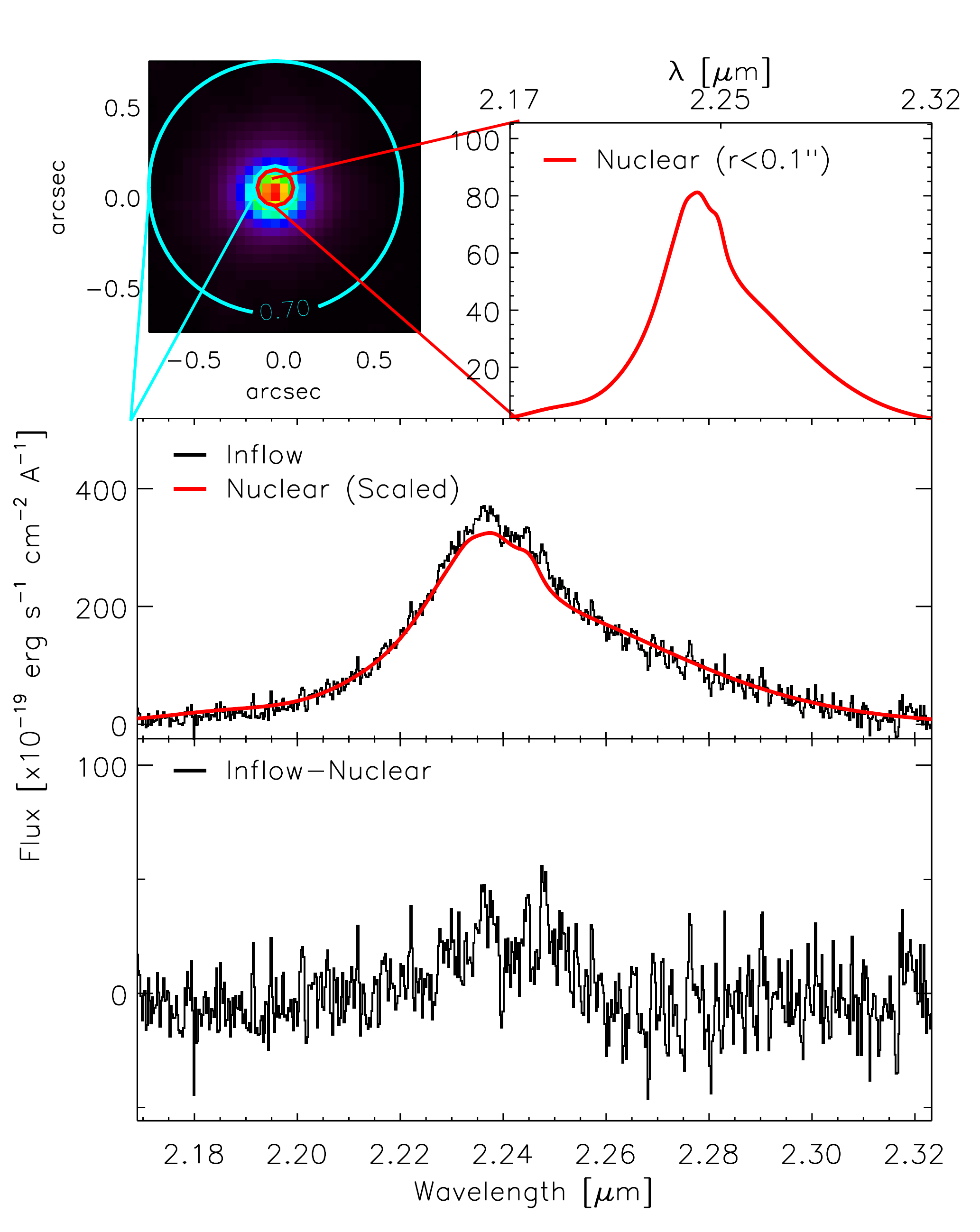}\\
	\caption{ Same as Figure~\ref{fig:outer_regions} but for the spectral
	region around H$\alpha$.}
	\label{fig:outer_regions_Ha}
\end{figure}
\begin{figure*}
	\centering

	\includegraphics[width=0.99\textwidth]{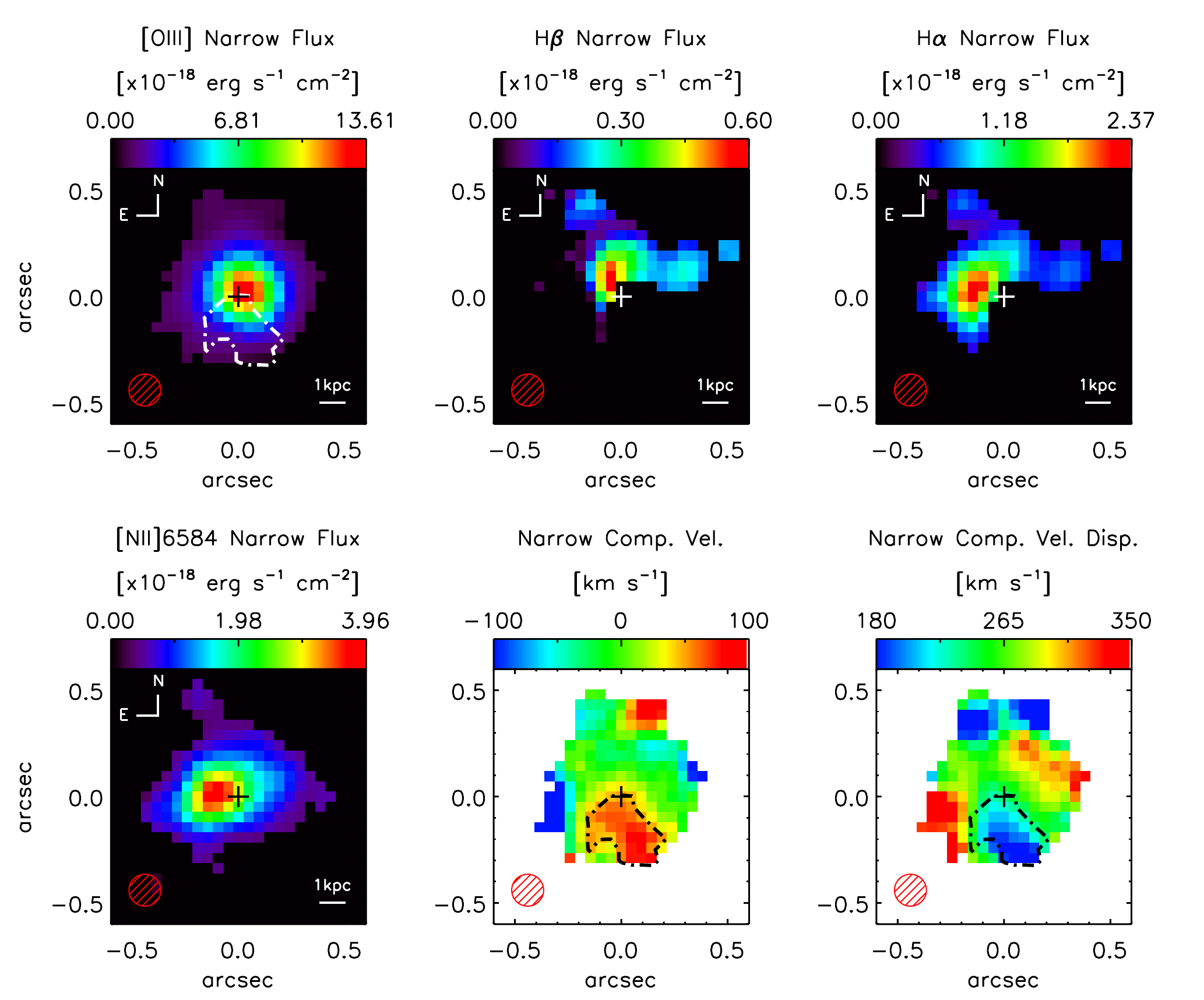}
	\caption{Maps of the narrow emission lines resulting from the spectral fitting. \textit{TOP}, from \textit{left to right} the panels show the fluxes of [\ion{O}{iii}] H$\upbeta$ and H$\alpha$. \textit{BOTTOM}, \textit{left}  flux of [\ion{N}{ii}]; \textit{center and right} velocity and velocity dispersion, which are linked to be the same for all components. In all maps, the cross marks the location of the continuum peak, the physical size is indicated by the 1~kpc scale and the size of the PSF is indicated by the red circle. The dotted contours on the [\ion{O}{iii}] flux, velocity and velocity dispersion maps indicate the inflow region from which the spectrum is extracted in Fig.~\ref{fig:inflow}. }
	\label{fig:kin_maps}
\end{figure*}

\section{Extended emission}

 Previous observations of 2QZJ0028, \citep[e.g.][]{canodiaz11,carniani15}, have reported evidence for extended nebular emission. Such previous studies have shown that the high velocity blueshifted [\ion{O}{iii}] emission traces a quasar driven outflow on scales of $3-5$~kpc, extending mostly towards the south east (SE), while narrow H$\upalpha$ and [\ion{O}{iii}] are mostly seen towards the north and to the west, tracing star formation in the host galaxy on scales of $3--5$~kpc.

We also investigate the extension of the nebular lines in our spectral cubes. As a first test for extended emission, we extract a
spectrum from the very central region ($r<0.1''$) and from a ring ($0.1''<r<0.7''$) sampling the outer regions. The extraction
regions are drawn in the top-left panel of Fig.~\ref{fig:outer_regions} on top of the continuum map. The nuclear spectrum is
fitted by using the usual components discussed above and the resulting fit is shown in Fig.~\ref{fig:outer_regions} top-right
restricted to the region around [\ion{O}{iii}]+H$\upbeta$ emission. The nuclear spectrum is then scaled to fit the outer spectrum,
by matching the intensity of only the broad wings of the Balmer lines, which are associated with BLR emission. Since the BLR is
certainly not resolved, such a component accounts for the emission that is apparently extended but actually associated with the
wings of the nuclear PSF. The subtraction of the scaled nuclear spectrum from the spectrum of the outer ring should have no
significant residuals if the bulk of the emission is unresolved. Instead the presence of significant residuals should highlight
extended and resolved emission. We find that such a procedure leaves only a very weak residual associated with the `intermediate'
components, which should trace the extended outflow. This implies that our high angular resolution observations are not sensitive
to the diffuse, very extended emission (detected in previous seeing-limited observations) associated with the extended outflow,
and that the intermediate blueshifted component is mostly dominated by the nuclear unresolved component of the outflow. However,
we do find a clear excess in the residuals associated with the narrow component. This is shown for the spectral region around
[\ion{O}{iii}] in the bottom panels of Fig.~\ref{fig:outer_regions}.
 Fig.~\ref{fig:outer_regions_Ha} shows the result of the same test for the spectral region around H$\alpha$+[NII], which
shows clear residuals of the intermediate/narrow components of these two nebular lines as well.

Using our resolved AO observations we can obtain 2D flux, velocity and velocity dispersion maps by fitting the individual spaxels with the same set of Gaussians described in Sect.~\ref{nuclear}. We fix the line width and velocity of both the broad components of H$\upalpha$ and H$\upbeta$ to the values obtained from the nuclear fit as these originate from the inner most few parsecs of the QSO which are unresolved and so do not vary spatially. We do the same for the H$\upalpha$ and H$\upbeta$ absorption components (i.e. fix to the values from the nuclear fit) and in addition tie the profile of the absorption components to the flux of the broad component, which essentially just allows the broad emission and absorption line profile to scale up and down. We also believe that we are not resolving the intermediate component of the emission lines, therefore we also fix the velocity and velocity dispersion to the values obtained from the fit to the inner nuclear region. In addition, the flux of the [\ion{O}{iii}]$\lambda$5007+$\lambda$4959 and H$\upbeta$ intermediate components are tied to scale with the H$\upbeta$ broad flux and similarly the flux of the H$\upalpha$ and [\ion{N}{ii}]$\lambda$6584+$\lambda$6548 intermediate components are tied to scale with the H$\upalpha$ broad line flux. This allows us to map all of the `narrow' components.  After fitting we then apply a S/N cut ($\rm S/N\approx3$) on the line fluxes to map each of the narrow emission line components.

\subsection{Indication for inflow}
Fig.~\ref{fig:kin_maps} shows the flux maps for the narrow component of the [\ion{O}{iii}]$\lambda$5007, H$\upbeta$, H$\upalpha$ and [\ion{N}{ii}]$\lambda$6584 emission lines. Fig.~\ref{fig:kin_maps} also shows the velocity and velocity dispersion, which we recall are tied for all narrow components.

The velocity field shows a hint of a blueshifted component with high velocity dispersion towards the South-East. This is probably a small part of the large scale outflow detected by previous studies in this direction, which is captured by our `narrow' component. Some indication of slightly blueshifted and slightly turbulent gas is detected towards the north-west; this may also be tracing an outflowing component, but not necessarily associated with the quasar as it may instead be associated with starburst winds due to star formation occurring in this region as found in previous studies and also suggested by our own data, as discussed in the next section.

\begin{figure}
	\centering
	\includegraphics[width=0.48\textwidth]{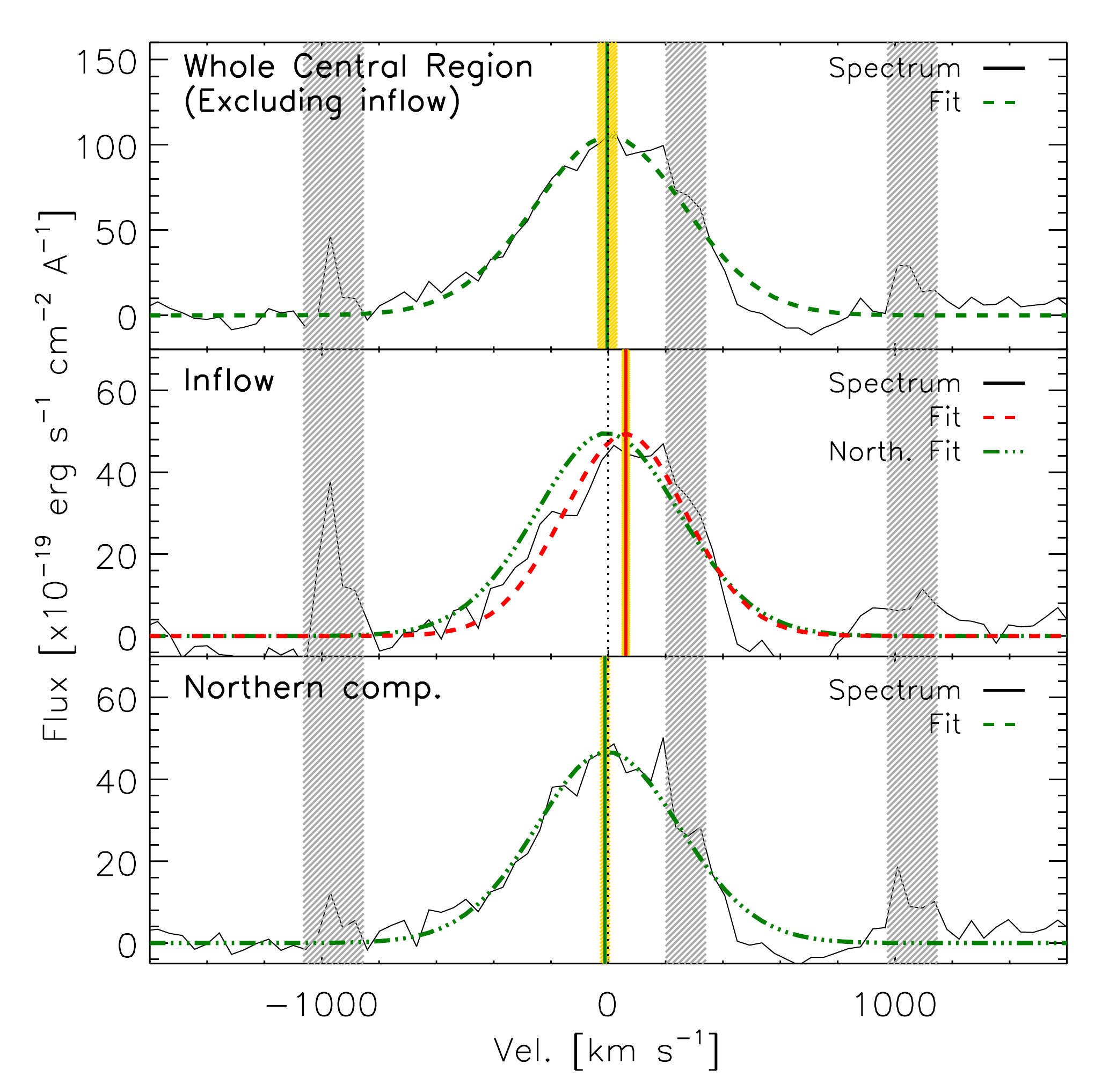}
	\caption{ \textit{Top}: Spectrum around the narrow [\ion{O}{iii}]$\lambda$5007 line extracted from the region of the galaxy excluding the inflow region . The green dashed line shows the fit to this narrow component and the green solid line indicates the resulting central velocity. \textit{Middle}: Spectrum extracted for the region of the galaxy where we see evidence
	for an inflow (i.e. with $v > 50 $~km~s$^{-1}$); the red dashed line shows the fit and the red solid lines indicates the resulting central velocity. \textit{Bottom}: Spectrum extracted for
	the northern region of the galaxy (i.e. direction opposite to the inflow); the green dashed line shows the fit and the green solid lines indicates the central velocity. The latter fit is also overplotted also in the central panel, to better illustrate the different profile and velocity shift. In all panels the (fixed) intermediate [\ion{O}{iii}] component has been subtracted. The grey dashed lines indicate the regions affected by strong OH sky lines. The yellow shaded region shows the uncertainties in the velocity shift for each component. }
	\label{fig:inflow}
\end{figure}

However, the most interesting feature of the kinematic map is a peculiar redshifted stream to the South-West. This could be
evidence for inflow of gas feeding the nuclear region and the QSO in particular.
 Indeed, in a scenario in which optical nebular emission from the opposite side of the galaxy (relative to our
line of sight) is obscured by dust in the host galaxy disc, the most likely interpretation of redshifted emission
is that it is tracing gas accreting on to the galaxy, likewise blueshifted emission is generally ascribed to outflowing gas.

 The velocity dispersion map seems to also suggest that the redshifted gas is also dynamically ``colder'', however,
as we shall see below, the significance of such a difference in velocity dispersion is low.


To  investigate further the putative inflowing component, Fig.~\ref{fig:inflow} shows the spectrum extracted from the inflow
location compared to that extracted from the rest of the central region, focusing only on the narrow [\ion{O}{iii}]$\lambda$5007
line (the intermediate component of [\ion{O}{iii}] has been subtracted from both spectra), which clearly shows that the inflowing
region has redshifted narrower emission compared to the main bulk of the central region. Note that the uncertainties in the
central velocities are indicated by the yellow shaded region, implying that the velocity shift of the `inflowing' region is indeed
significant compared to the rest of the emission.

We also extract the spectrum from a region on the opposite side of the inflow (i.e. towards the NE), which is shown in the bottom panel of Fig.~\ref{fig:inflow}. This illustrates that the northern region does not show evidence for a symmetric blueshifted velocity, which one would expect in the case that the inflow is actually part of a rotating system. A rotating disc is however excluded by the global velocity field, which is quite different than expected from any rotation pattern.

 The quantitative comparison of the velocity and velocity dispersions of the various components
is reported in Table~\ref{tab:velpars}. The velocity of the ``inflowing'' component differs from the northern region
by 3.5$\sigma$ and differs from the rest of the emitting region by 2.7$\sigma$. As mentioned above, the differences in terms
of velocity dispersion are more marginal, at the level of $\sim$2$\sigma$.

\begin{table}
	\caption{Velocity and velocity dispersion shifts of the narrow  [\ion{O}{iii}]$\lambda$5007 shown in Fig.~\ref{fig:inflow} extracted from different regions within the galaxy. }
	\label{tab:velpars} 
	\begin{tabular}{@{}lcc@{}}  
	\hline
	Region & Velocity $\rm [km~s^{-1}]$& Dispersion $\rm [km~s^{-1}]$ \\
	\hline
	Whole Region & $-4 \pm 20 $ & $260 \pm 19 $\\
	\ \ (excl. inflow) & & \\
	Inflow & $61 \pm 13 $ & $204 \pm 14 $\\
	Northern Comp. & $-12 \pm 16 $ & $240 \pm 16 $\\
	\hline
	\end{tabular}\\
\end{table}

We shall note that such inflowing feature was not identified in previous seeing-limited data, because the stronger [\ion{O}{iii}] emission from the rest of the galaxy smears away the inflowing signature. Indeed, we have verified that by smoothing our cube to the angular resolution of previous data, and by extracting the velocity and velocity dispersion as in previous works (i.e. by extracting the first and second moments of the [\ion{O}{iii}] line without separating narrow and intermediate components), the signature of the inflowing stream disappears and we obtain kinematic maps consistent with previous studies.

 We shall note that although the angular resolution of these observations are unprecedented for this object, it is not possible to exclude other scenarios in which the redshifted feature could be associated with an infalling gas rich companion galaxy, although as we will discuss in Sect.~\ref{sect:bpt} the observed line ratios are not consistent with a star forming system.

\begin{figure*}
	\centering
	\includegraphics[width=0.8\textwidth]{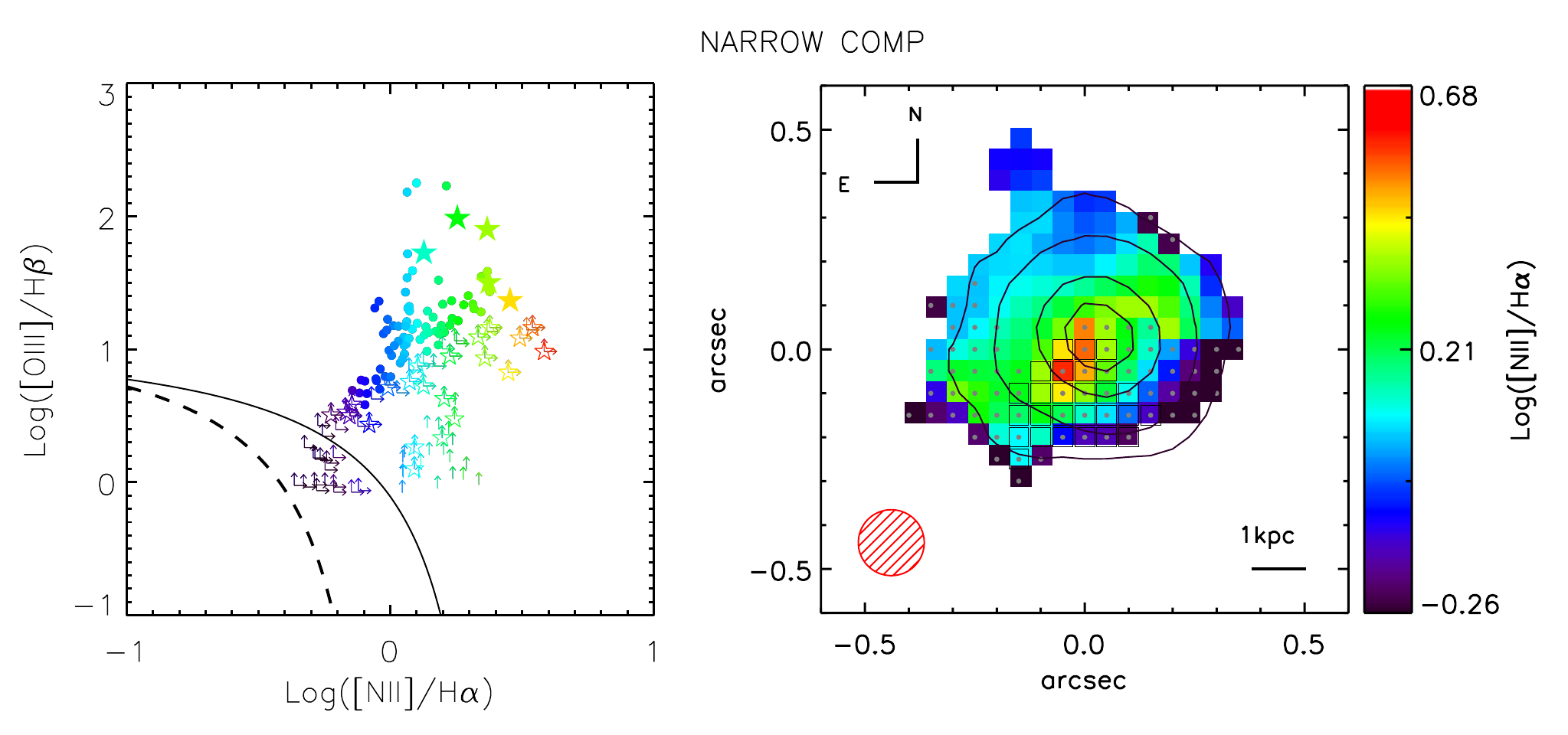}
	\caption{\textit{Left} BPT diagnostic diagram for the narrow component of the emission, colour-coded by the ratio of
	[\ion{N}{ii}]/H$\upalpha$. Stars indicate regions associated with the putative inflowing region. \textit{Right} Map of the galaxy where each pixel is colour-coded	by the corresponding [\ion{N}{ii}]/H$\upalpha$ ratio, therefore indicating the regions closer to the demarcation line and those clearly dominated by AGN ionization. We use upper limits on the H$\upalpha$ and/ or H$\upbeta$ fluxes, where they are not detected, which places lower limits on the BPT ratios as shown by the arrows in the left plot and these regions are also indicated on the map (\textit{right}) with a grey dot. Note that the physical size is indicated by the 1 kpc scale and the size of the PSF is indicated by the red circle}
	\label{fig:bpt}
\end{figure*}

\subsection{Inflow rate}
{\it Assuming our preferred interpretation of the redshifted region, as tracing a stream of inflowing gas,}
it is possible to infer an inflow rate, although this requires some assumptions. The [\ion{O}{iii}] luminosity of the gas associated with the streaming region is about $\rm 10^{43}~erg~s^{-1}$. If one assumes an electron density of the line emitting clouds of about $\rm 10^2~cm^{-3}$, which is rather typical of ionized gas in galactic discs \citep[e.g.][]{belfiore15}, the [\ion{O}{iii}] luminosity implies a mass of ionized gas of $M_{\rm i} \sim 5 \times10^7~\rm M_{\odot}$ \citep[using Eq. 1 in ][]{canodiaz11}. However, \cite{carniani15} have shown that this is actually a lower limit since it does not fully account for the mass of gas in lower ionization stages than that traced by [\ion{O}{iii}]; the mass is actually at least a factor of two higher, i.e. $M_{\rm i} \sim 1 \times10^8~ \rm M_{\odot}$, especially if the gas is in the LINER-like ionization region of the BPT diagram, as discussed below. The gas density averaged along the streaming structure is given by $\langle \rho_{\rm i} \rangle = M_{\rm i}/(A~l~(\sin{\theta})^{-1})$, where $A$ is the cross section of the streaming structure perpendicular to the streaming direction, $l$ is the {\it observed} length of the streaming structure ($\sim 2$~kpc) and $\theta$ is its inclination angle relative to the line of sight ($\theta=90^{\circ}$ means perpendicular to the line of sight), which takes into account that we see the streaming structure only in projection. Note that the previous average also takes into account the average filling factor of gas clouds.

The inflow rate is then given by $\dot{M}_{\rm i} = \langle \rho_{\rm i} \rangle~A~v~(\cos{\theta})^{-1} = M_{\rm i}~\frac{v}{l}~\tan{\theta}$, where $v$ is the {\it observed} velocity of the stream ($\rm \sim 100~km~s^{-1}$) and the term $\rm (\cos{\theta})^{-1}$ takes into account that we see such velocity in projection. The main source of uncertainty is the term $\tan{\theta}$, which is unknown. However, we can provide some constraints on $\theta$ by taking into consideration that the actual (de-projected) extent of the streaming structure is unlikely to be larger than the typical galactic scales, i.e. $<$10~kpc, which implies $\theta >25^{\circ}$, and that the actual (de-projected) inflow velocity of the stream cannot be much larger than the maximum infall velocity observed in simulations in even the most massive systems, i.e. $<$600~km~s$^{-1}$ \citep{costa15}, which implies $\theta < 80 ^{\circ}$. Therefore the inflow rate must be in the range: $3<\dot{M}_{\rm i}/(\rm M_{\odot}~yr^{-1})<33$.

 This inflow rate is unlikely enough to fuel the black hole and maintain 
the observed luminosity. Indeed for this quasar $\lambda L_{5100}=3
\times10^{46}~\rm erg~s^{-1}$, implying $L_{\rm bol}\sim 2 \times10^{47}~\rm
erg~s^{-1}$ assuming the bolometric correction given in \cite{marconi14}, which
requires an accretion rate of $\rm 35~M_{\odot}~yr^{-1}$, assuming a radiative
efficiency of 0.1.  This suggests that the quasar phase is a transient
phenomenon, as expected by cosmological simulations.

We note that also the outflow rate inferred in the previous section is actually higher than the inflow rate inferred here. If confirmed, this would imply that the outflow must be a transient phenomenon. However, it is more likely that the discrepancy is mostly driven by the large uncertainties in both estimates.

One should notice that the inflowing gas can also deliver fuel for star formation in the central region of the galaxy, which can contribute to the early formation of the galactic bulge.

Such major inflow of gas could result either by disc instabilities, a non-axysimmetric potential (a stellar bar) or could have been induced by the tidal interaction with a companion galaxy.

\subsection{Gas excitation map}
\label{sect:bpt}
To explore the ionization source of the emission, we use the line ratio diagnostics [\ion{O}{iii}]/H$\upbeta$ versus.
[\ion{N}{ii}]/H$\upalpha$, i.e. one of the so called BPT diagrams. The intermediate component, tracing the central unresolved
outflow has line ratios typical of the Narrow Line Region (NLR) in AGNs ($\log([\ion{O}{iii}]$/H$\upbeta$) = 1.17,
$\log([\ion{N}{ii}]$/H$\upalpha$) = 0.13). Regarding the narrow components, Fig.~\ref{fig:bpt} shows where each spaxel lies on the
BPT diagram in relation to the demarcation line between AGN and star-forming emission, which is subsequently colour-coded by the
[\ion{N}{ii}]/H$\upalpha$ ratio. Spaxels marked with a dot are regions with lower limit on one or both line ratios. The line
ratios are consistent with AGN ionization across most of the sampled regions.
 The errorbars are not shown to avoid overcrowding and for sake of clarity, however errorbars range from 0.1 to 0.4 dex (and
lower limits are at 2$\sigma$), hence a fraction of the region could still be consistent with star formation within uncertainties.

There is an indication that the inner regions are subject to stronger AGN ionization, while the regions extending towards the north and west tend to approach the \ion{H}{ii} star-forming location on the BPT diagram, although still within the AGN demarcation. This result connects nicely with the finding of \cite{canodiaz11}, who trace star formation in the form of narrow H$\upalpha$ emission in the outer regions towards the north and east; the latter are however missed by our AO data, so the regions sampled by us never reach the \ion{H}{ii} locus on the BPT diagram, although the trend suggests an increasing contribution from star formation relative to AGN ionization.

Towards the SE \cite[the direction of the outflow identified by][]{canodiaz11,carniani15} H$\upbeta$ is not detected, but the [\ion{N}{ii}]/H$\upalpha$ ratios and the lower limits on [\ion{O}{iii}]/H$\upbeta$ are consistent with AGN photoionization and also with strong shocks \citep{allen08}.

Unfortunately, along the inflowing streams both H$\upalpha$ and H$\upbeta$ are undetected so we only have lower limits on both [\ion{N}{ii}]/H$\upalpha$ and [\ion{O}{iii}]/H$\upbeta$. The values are not very constraining. Here we only note that the line ratios are consistent with both some minor contribution from star formation and also with shock excitation \cite{allen08}. More data are required to better constrain the nebular line diagnostics in this region.


\section{Conclusions}
We have presented new near-IR integral field spectroscopic observations of the QSO 2QZJ0028 at $z=2.4$ by using SINFONI together with its Adaptive Optics module, which enables us to achieve an angular resolution ($\sim 0.15$~arcsec) close to the diffraction limit of the VLT. 2QZJ0028 is the first system in which previous observations had found direct evidence for a quasar-driven outflow quenching star formation in its host galaxy. The new data have enabled us to investigate in detail the physics of this system.

The analysis of the nuclear spectrum has enabled the following findings:
\begin{itemize}
\item The nuclear spectrum reveals the presence of broad, blueshifted H$\upbeta$ absorption, which is tracing outflowing gas along the line of sight with velocities up to $10,000$~km~s$^{-1}$.
\item Population of the  $n=2$ hydrogen level, required to see H$\upbeta$ in absorption, requires large gas densities. We have provided some simple photoionization models quantitatively confirming that gas densities $\rm > 10^8\--10^9~cm^{-3}$ are required to explain the observed H$\upbeta$ absorption depth. The implied total column densities of hydrogen are in the range $10^{22}<N_{\rm H}<10^{24}~\rm cm^{-2}$.
\item The H$\upalpha$ line does not show evidence for absorption, very likely because it is dominated by the much stronger (relative to H$\upbeta$) emission component. However, the H$\upalpha$ emission line profile does show a clear asymmetry, which can be nicely modelled in terms of the blue shoulder being suppressed by absorption associated with the dense outflowing gas.
\item The relative depth of the absorption inferred for H$\upbeta$ and H$\upalpha$ clearly reveals that the absorption features must be saturated and that the outflowing gas must cover the background radiation (the accretion disc) only partially. By comparing the H$\upbeta$ and H$\upalpha$ profile we can infer the covering factor of the outflowing clouds. We find that the covering factor varies significantly as a function of the gas velocity, between 0.1 and about 0.7. This implies an upper limit on the size of the outflowing clouds of about $\rm 10^{-3}$ pc.
\item The properties of the outflowing absorbing gas are consistent with the scenario in which the outflowing clouds originate from instabilities in the accretion disc and/or from the BLR.
\item Outflow rate and kinetic power of this nuclear outflow are difficult to estimate due to the large uncertainties on the properties of the outflow gas. However, order of magnitude estimates suggest that the kinetic energy of the nuclear outflow is comparable to the kinetic energy of the outflow detected on large scale. This finding is consistent with energy-conserving models of quasar-driven outflows.
\item Assessing the fraction of quasars with Balmer broad and blueshifted absorption lines is not simple due to technical difficulties in identifying the Balmer lines in large samples of high redshift quasars (in particular the need of large number of broad band spectra with high S/N), and also due to the difficulties in disentangling Balmer absorption in the complex spectra of quasars. However, we note that the asymmetric profile of the Balmer lines seen in several quasar may be ascribed to absorption associated with nuclear dense outflows as in 2QZJ0028. More generally, our results prompt for a fresh re-analysis of the complex quasar spectra, by allowing for the potential presence of Balmer absorption.
\item We also note that the absorption of the blue side of the Balmer lines due to such dense outflowing gas has a significant impact on the determination of black hole masses based on virial estimators, prompting for a revision of the black hole masses in quasars with asymmetric lines.
\end{itemize}

The analysis of the spatially extended component of the nebular emission lines is limited to within the central arcsecond since our high resolution data are not sensitive to diffuse emission on larger scales. In particular, while we find some hints of the extended outflow traced by previous observations, the bulk of the diffuse outflow is not detected (although we do detect a intermediate [\ion{O}{iii}] component associated with the inner part of the outflow). However, the analysis of the extended nebular emission has enabled the following interesting findings:

\begin{itemize}
\item The kinematic analysis reveal  a region of redshifted nebular lines,  pointing towards the centre, detected at
the $\sim 3\sigma$ level relative to the surrounding regions. We suggest
that this could be tracing a stream of inflowing gas, although other scenarios cannot be excluded.
\item  In the inflow scenario,  the gas inflow rate is inferred to be in
the range of $\rm 3 - 33~M_{\odot}~yr^{-1}$.
 This inflow rate is unlikely to account for the large accretion rate
inferred for this very luminous quasar, implying that this is likely a
transient phase (at least in this extremely luminous stage).
\item We could also constrain the gas excitation mechanism in the central region, by locating the different central regions on the
classical BPT diagnostic diagram. Results show that the central region is mostly dominated by AGN excitation, while the gas in the outer 1-2 kpc tend to move towards the \ion{H}{ii}/star-forming region of the BPT diagram, consistent with previous results. The region associated with the streaming inflow is difficult to characterize, since the narrow component of H$\upbeta$ is undetected. However, limits on the BPT diagram suggests that the gas in this region could be predominantly
excited by shocks.
\end{itemize}

\section{Acknowledgements}
Y.K. acknowledges support from grant DGAPA PAIIPIT IN104215 and CONACYT grant168519.
R.M. and R.W. acknowledge support from the United Kingdom Science Facilities Council, through grant ST/M001172/1.
R.M. acknowledges support by the Science and Technology Facilities Council (STFC) and the ERC Advanced Grant 695671. 
``QUENCH''.
The data presented in this paper can be accessed at the ESO web archive: http://archive.eso.org/eso/eso\_archive\_main.html.
This work is based on observations made with ESO telescopes at the La Silla Paranal Observatory under programme ID 090.A-0282(A).

\bibliographystyle{mnras}
\setlength{\labelwidth}{0pt}
\bibliography{myrefs}

\appendix
\label{appendix}

\section[Lack of \ion{Fe}{ii} contribution and implications]{Lack of \ion{Fe}{ii} contribution and implications}
\label{sect:feii}

A possible concern of the detection of H$\upbeta$ absorption is that it could actually be ascribed to a region of minimum emission between the H$\upbeta$ emission and a \ion{Fe}{ii} broad bump. As already mentioned in Sect.~\ref{nuclear}, even if including an \ion{Fe}{ii} component is included, this is set to zero by the fit. In this appendix we perform some additional tests that further rule out the presence
of significant \ion{Fe}{ii} emission, and we also discuss some more broader potential implications for the interpretation of quasar spectra.

\begin{figure}
	\centering
	\includegraphics[width=0.48\textwidth]{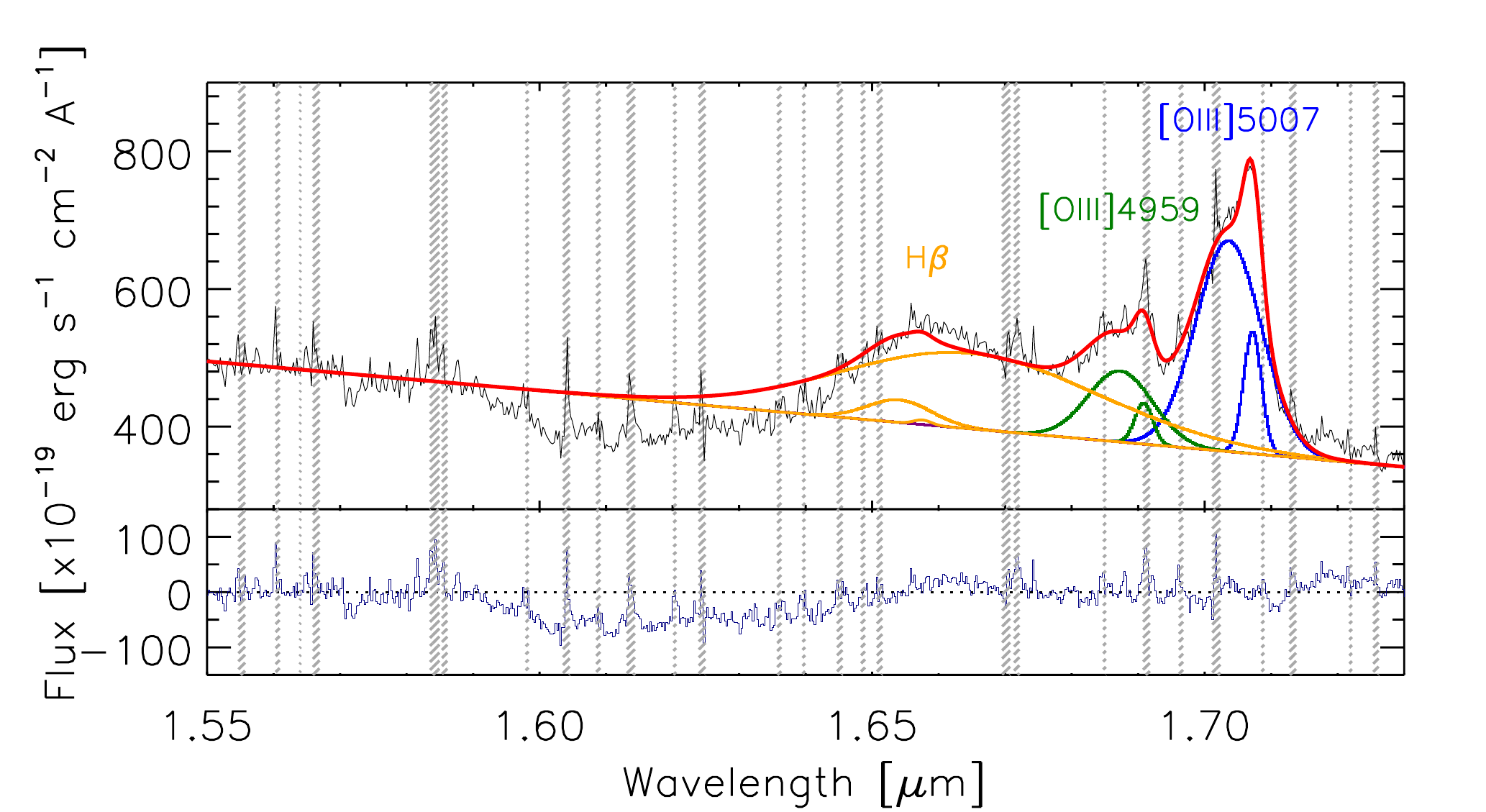}
	\caption{Spectrum extracted from the central arcsec in the \textit{H}-band. Here we have attempted to fit the spectrum with pure emission, i.e. have removed the absorbing components from the
	fit. Different colours indicate different emission components, more specifically: [\ion{O}{iii}]$\lambda$5007 (blue), [\ion{O}{iii}]$\lambda$4959 (green), H$\upbeta$ emission (orange). The total
	resulting fit is shown by the red line. The residuals from the fit are shown in the bottom panel. It can been seen that the fit does not require a significant \ion{Fe}{ii} contribution and
	that the fit requires an absorption component to account for the strong negative residual. The grey dashed lines indicate the regions affected by strong OH sky lines.}
	\label{fig:no_abs}
\end{figure}

Fig.~\ref{fig:no_abs} shows the spectrum extracted from the central arcsec in the \textit{H}-band. In this case we allow the fit to use {\it only emission} components, including \ion{Fe}{ii}, on top of the continuum. The resulting best fit does not attempt to recover the dip at $\sim 1.61~\upmu \rm m$ (blueward of H$\upbeta$) by introducing \ion{Fe}{ii} emission. The latter component is still set to zero.
Remarkably, the fit shows a clear strong negative residual, which cannot be accounted for by any combination of continuum normalization/slope and emission features. In order to account for this residual negative feature an additional absorption component must be included, as we have done in the paper.

\begin{figure}
	\centering
	\includegraphics[width=0.48\textwidth]{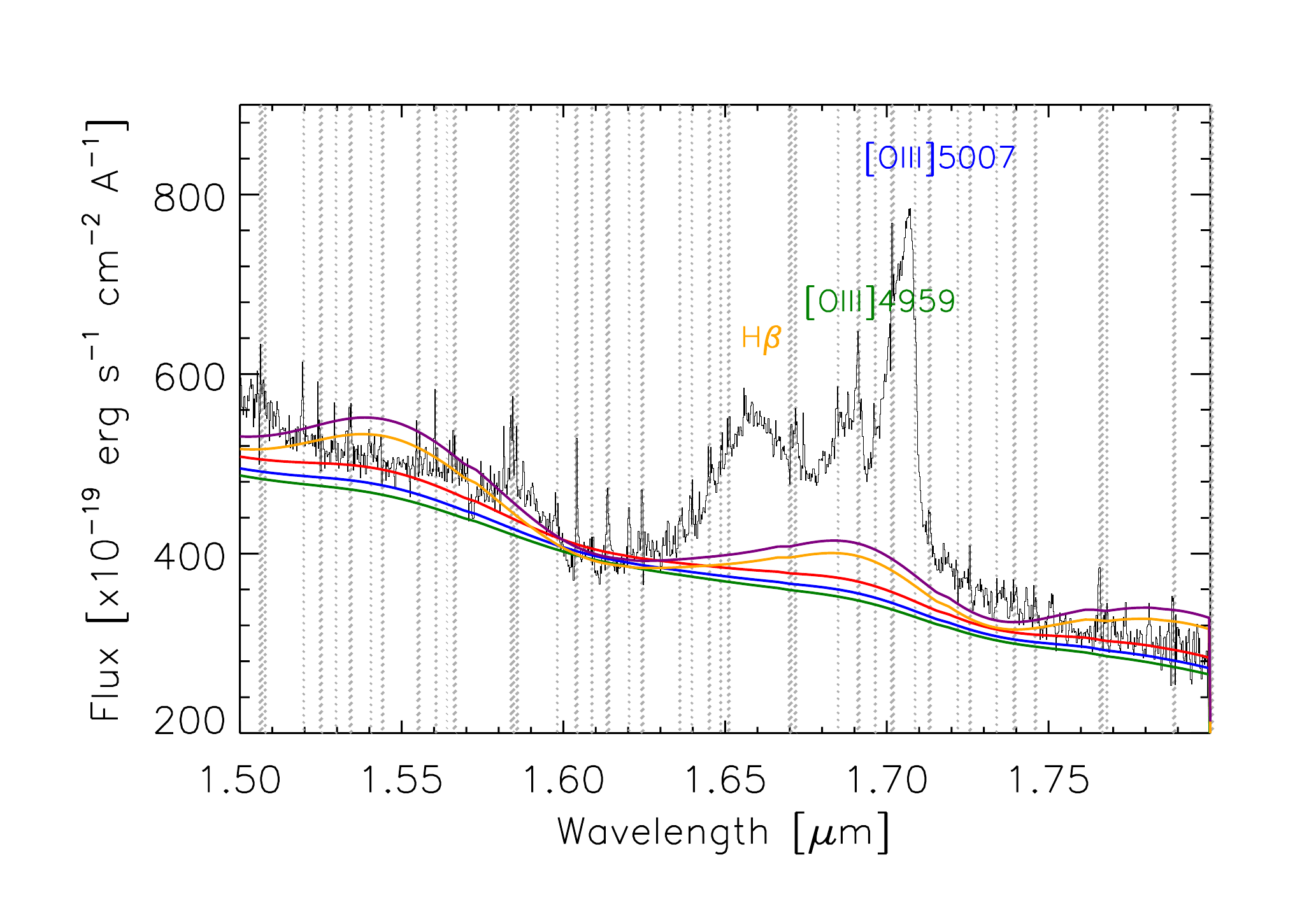}
	\caption{Spectrum extracted from the central arcsec showing the \textit{H}-band region of the spectrum. Here we have forced the introduction of an \ion{Fe}{ii} template, on top of a continuum
	passing through the dip of the absorption feature. The \ion{Fe}{ii} template is smoothed to the width of H$\upbeta$ and then scaled by various different factors. Clearly none of the \ion{Fe}{ii} scalings can match the relatively featureless spectrum blueward of the H$\upbeta$ absorption and redward of the [\ion{O}{iii}] emission line.}
	\label{fig:feii}
\end{figure}

In Fig.~\ref{fig:feii}, we attempt to force an \ion{Fe}{ii} component to the fit. We start with the continuum slope given by the previous attempt (the result does not change by using other slopes) and normalize it to the continuum at the bottom of the absorption dip (i.e. we are now hypothetically assuming that the dip of the absorption feature is the true continuum level). Line of different colours indicate the effect of adding the contribution of the \ion{Fe}{ii} template, smoothed to the same velocity dispersion as H$\upbeta$, with a range of normalization factors. Clearly, it is not possible for the \ion{Fe}{ii} component to reproduce the observed spectrum outside the dip. The main issue is that, outside the `dip' and outside the H$\upbeta$+[\ion{O}{iii}] emission group, the spectrum is essentially featureless, with a smooth slope.Such a featureless continuum cannot be reproduced by any \ion{Fe}{ii} emission, which would imply a `bumpy' spectrum.

Within this context we note that spectra with narrower spectral coverage (either because of redshift or because of instrumental capabilities) may not have the same capability of discriminating between a `dip' associated with H$\upbeta$ blueshifted absorption and \ion{Fe}{ii} emission. In particular, we suspect that a number of quasar spectra, with narrow spectral coverage, showing a depression blueward of H$\upbeta$, might have been erroneously fitted with \ion{Fe}{ii} emission, while the depression may actually be associated with broad blueshifted H$\upbeta$ absorption.

Even if \ion{Fe}{ii} emission is present, we suspect that the lack of \ion{Fe}{ii} emission in the spectral region just blueward of H$\upbeta$ might still be due to H$\upbeta$ blueshifted absorption. A clear hint of this potential issue is seen in the results obtained by \cite{zhang15}: in the spectrum of this quasar Balmer absorption, and in particular broad blueshifted H$\upalpha$ absorption is unambiguously detected; the corresponding H$\upbeta$ absorption is located at exactly the position of the putative `\ion{Fe}{ii}' typically observed in quasar spectra just bluewards of H$\upbeta$. If confirmed with additional analyses and observations, this would imply that the \ion{Fe}{ii} quasar `templates' themselves might be affected by this issue and, therefore, their use may force the misidentification H$\upbeta$ blueshifted absorption. 

These are all scenarios that need to be tested further with more detailed modelling and also with new, high S/N broad band spectra. More generally, all of these considerations prompt for a re-analysis of quasar spectra with a fresh look, by including the possibility of Balmer absorption.

\section{Previous SINFONI Observations}
\label{sect:previous}
\begin{figure}
	\centering
	\includegraphics[width=0.45\textwidth]{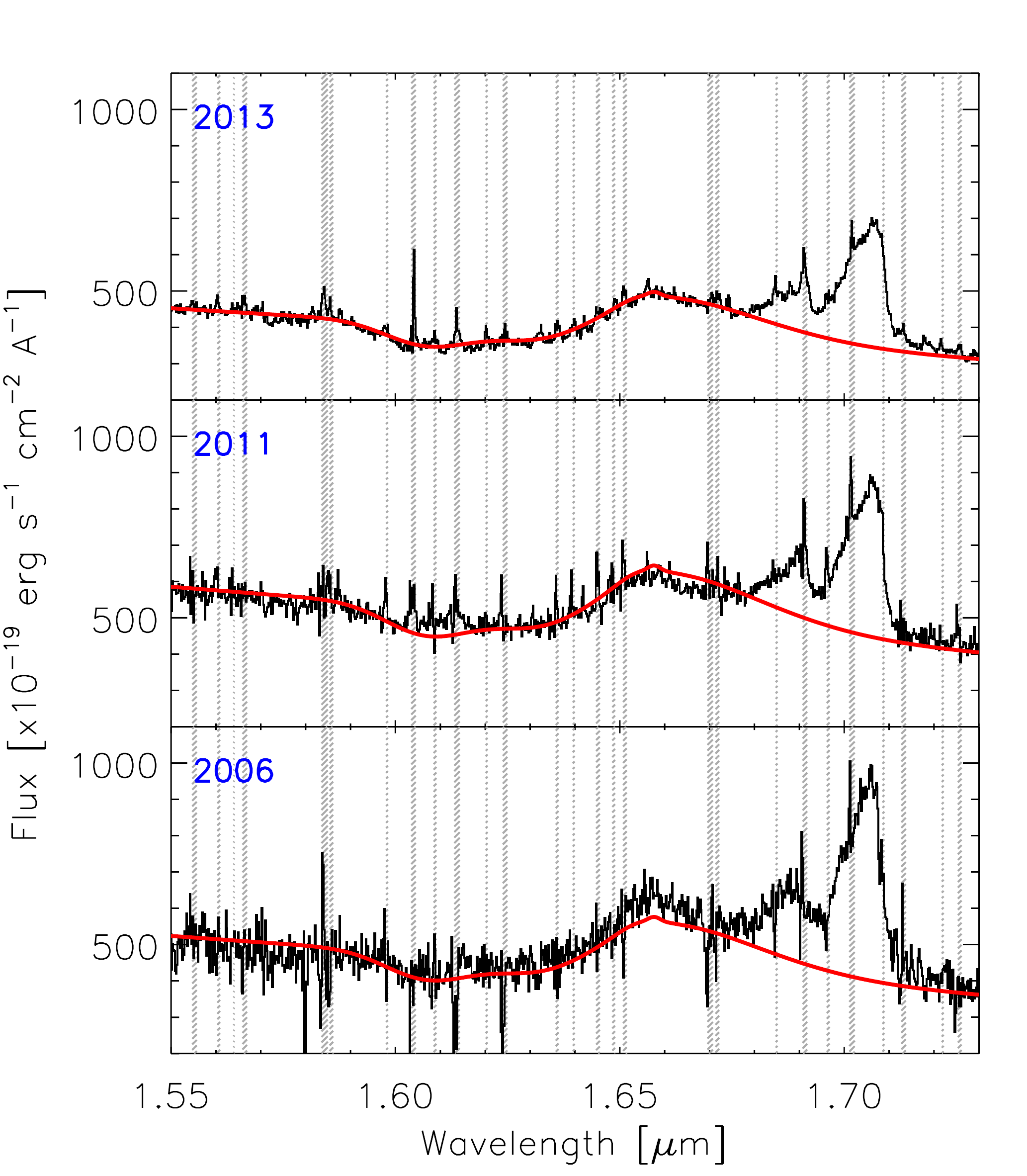}
	\caption{SINFONI \textit{H}-band spectra of 2QZJ0028 taken at the three epochs. \textit{Top:} this work, 2013 (AO-assisted), \textit{Middle:} 2011 (seeing-limited) from \citet{carniani15} and  \textit{Bottom:} 2006 (seeing-limited) from \citet{canodiaz11}. All spectra are extracted from a circular aperture twice the size of the seeing in each observation around the centre of the galaxy. The best fit to our data (scaled to match the continuum of each observation) is shown in red indicating that the previous	observations are consistent with the same absorption feature. The grey dashed lines indicate the regions affected by strong OH sky lines.}
	\label{fig:3epochs}
\end{figure}

In this appendix we compare the \textit{H}-band spectrum obtained from our data with the SINFONI spectra of the same quasar obtained in two previous epochs. Fig.~\ref{fig:3epochs} show the SINFONI spectra obtained in the three epochs. In red we have over-plotted the best fit to our model (from Fig.~\ref{fig:fitting}), scaled to match the continuum of each of the two previous observations. Note that the [\ion{O}{iii}] lines are ignored, as these can vary due to aperture effects. It can be seen that both the 2006 and 2011 data are generally consistent (within the noise) with a H$\upbeta$ absorption feature. Yet, some weak indication of possible variability may be seen both in the absorption and emission of H$\upbeta$.

\label{lastpage}
\end{document}